\documentclass{emulateapj}
\usepackage{apjfonts}
\usepackage{graphicx}
\usepackage{xspace}
\usepackage{booktabs,longtable,tabu}

\begin{document}

\lefthead{{\em NuSTAR} Serendipitous Sources with {\em Chandra}}
\righthead{Tomsick et al.}

\submitted{Accepted by ApJ}

\def\lsim{\mathrel{\lower .85ex\hbox{\rlap{$\sim$}\raise
.95ex\hbox{$<$} }}}
\def\gsim{\mathrel{\lower .80ex\hbox{\rlap{$\sim$}\raise
.90ex\hbox{$>$} }}}

\title{{\em Chandra} Observations of {\em NuSTAR} Serendipitous Sources near the Galactic Plane}

\author{John A. Tomsick\altaffilmark{1}, 
George B. Lansbury\altaffilmark{2},
Farid Rahoui\altaffilmark{3}, 
James Aird\altaffilmark{2},
David M. Alexander\altaffilmark{4},
Ma\"ica Clavel\altaffilmark{5},
AnaSofija Cuturilo\altaffilmark{6},
Francesca M. Fornasini\altaffilmark{7}, 
JaeSub Hong\altaffilmark{7}, 
Lizelke Klindt\altaffilmark{4},
Daniel Stern\altaffilmark{8}
}

\altaffiltext{1}{Space Sciences Laboratory, 7 Gauss Way, University of California, 
Berkeley, CA 94720-7450, USA}

\altaffiltext{2}{Institute of Astronomy, University of Cambridge, Madingley Road, 
Cambridge CB3 0HA, UK}

\altaffiltext{3}{Department of Astronomy, Harvard University, 60 Garden Street, 
Cambridge, MA 02138, USA}

\altaffiltext{4}{Centre for Extragalactic Astronomy, Department of Physics, University 
of Durham, South Road, Durham DH1 3LE, UK}

\altaffiltext{5}{Univ. Grenoble Alpes, CNRS, IPAG, F-38000 Grenoble, France}

\altaffiltext{6}{Seattle Pacific University, 3307 3rd Ave West, Seattle, WA 98119-1997, USA}

\altaffiltext{7}{Harvard-Smithsonian Center for Astrophysics, 60 Garden Street,
Cambridge, MA 02138, USA}

\altaffiltext{8}{Jet Propulsion Laboratory, California Institute of Technology, 
4800 Oak Grove Drive, Pasadena, CA 91109, USA}

\begin{abstract}

The {\em NuSTAR} serendipitous survey has already uncovered a large number of 
Active Galactic Nuclei (AGN), providing new information about the composition of 
the Cosmic X-ray Background.  For the AGN off the Galactic plane, it has been
possible to use the existing X-ray archival data to improve source localizations, 
identify optical counterparts, and classify the AGN with optical spectroscopy.
However, near the Galactic Plane, better X-ray positions are necessary to 
achieve optical or near-IR identifications due to the higher levels of source
crowding.  Thus, we have used observations with the {\em Chandra X-ray
Observatory} to obtain the best possible X-ray positions.  With eight observations,
we have obtained coverage for 19 {\em NuSTAR} serendips within $12^{\circ}$ of
the plane.  One or two {\em Chandra} sources are detected within the error circle
of 15 of the serendips, and we report on these sources and search for optical
counterparts.  For one source (NuSTAR~J202421+3350.9), we obtained a new optical
spectrum and detected the presence of hydrogen emission lines.  The source is
Galactic, and we argue that it is likely a Cataclysmic Variable.  For the other
sources, the {\em Chandra} positions will enable future classifications in order
to place limits on faint Galactic populations, including high-mass X-ray binaries
and magnetars.

\end{abstract}

\keywords{surveys --- stars: white dwarfs --- stars: neutron ---
  stars: black holes --- X-rays: stars}

\section{Introduction}

With its ability to focus hard X-rays, the {\em Nuclear Spectroscopic Telescope
Array (NuSTAR)} provides unprecedented sensitivity above $\sim$10\,keV
\cite{harrison13}.  Thus, surveys with {\em NuSTAR} allow us to study faint 
populations of high-energy sources, including AGN as 
well as Galactic populations such as X-ray binaries, Cataclysmic Variables (CVs), 
pulsar wind nebulae, supernova remnants, and stars with active coronae.  The 
{\em International Gamma-ray Astrophysics Laboratory (INTEGRAL)} satellite and 
the Burst Alert Telescope (BAT) on the {\em Swift} satellite have surveyed the 
sky at 17--100\,keV and 15--55\,keV, respectively \citep{ajello12,bird16}, but 
{\em NuSTAR} is extending to flux levels that are approximately 2 orders of 
magnitude lower.  For high-mass X-ray binaries (HMXBs), {\em INTEGRAL} has been 
used to constrain their surface density ($\log{N}$-$\log{S}$) down to 
$\sim$$10^{-11}$\,erg\,cm$^{-2}$\,s$^{-1}$ \citep{lutovinov13}, but \cite{tomsick17} 
demonstrates the possibility of extending the constraints down below 
$10^{-13}$\,erg\,cm$^{-2}$\,s$^{-1}$ with {\em NuSTAR}.  

Using {\em NuSTAR} data from the first 40 months of the mission, Lansbury 
et al.~(2017, henceforth L17)\nocite{lansbury17} carried out a search for 
serendipitously detected {\em NuSTAR} sources (i.e., serendips).  L17 compiled
a catalog of 497 sources in the primary source catalog and 64 sources in the 
secondary source catalog.  As described in L17, the secondary catalog consists
of sources that are robustly detected using a secondary source detection method 
that is different from the method used for the primary catalog.  The 3--24\,keV 
energy band was used for the serendipitous survey, and the sky coverage is 
$\sim$13\,deg$^{2}$.  Of the 561 sources in both the primary and secondary
catalogs, optical identifications have been obtained for 318.  Optical spectroscopy
shows that 297 are likely AGN and 21 are likely Galactic, but the identify of
the optical counterpart is uncertaint for five of these.  The nature of 16 of the
Galactic sources with secure counterpart identifications in the primary and secondary
catalogs has been investigated \citep{tomsick17}, and they include stars, CVs,
low-mass X-ray binaries (LMXBs), and HMXBs.  In addition, at least one previously
known magnetar is also among the serendips.  In fact, more magnetars may be 
present among the unclassified serendips; however, given that most of 
the serendip classifications thus far have been based on optical spectra, the 
magnetars, which are typically very faint in the optical, would not have been 
found.

\cite{tomsick17} investigated how the completeness of source classifications
depends on Galactic latitude, and while the completeness is 63\% for sources 
in the primary catalog that are more than $10^{\circ}$ from the Galactic 
plane, this drops to 32\% at $5^{\circ}$--$10^{\circ}$, and only 7 of 57 (12\%) 
of sources have been classified at $|b|$$<$$5^{\circ}$.  A major reason 
for the incompleteness at low Galactic latitudes is source confusion.  L17 
searched for X-ray counterparts for {\em NuSTAR} serendips with coverage by 
{\em XMM-Newton}, {\em Swift}, or {\em Chandra}, and used the X-ray positions 
to search for optical or near-IR (OIR) matches.  L17 used the separations between 
the X-ray and OIR positions along with the sky density of OIR sources to estimate 
the spurious matching fractions.  For high-latitude sources ($|b|$$>$$10^{\circ}$) 
with {\em XMM} or {\em Swift} positions, the spurious matching fractions were 
6--16\%, but they were 1.2--1.7\% for high-latitude sources with {\em Chandra} 
positions.  Thus, the reliability of the L17 classifications is very high for 
the $|b|$$>$$10^{\circ}$ serendips.  Although L17 did not give specific numbers
for low-latitude sources, the larger OIR sky densities would naturally lead
to significantly larger spurious matching fractions.  Thus, we have obtained 
{\em Chandra} observations for low-latitude sources to improve the X-ray 
positions and find OIR counterparts.

Here, we report on {\em Chandra} observations obtained to attempt to identify and
classify more of the {\em NuSTAR} serendips close to the Galactic plane.  In
Section~2, we describe the {\em Chandra} observations that we use and our analysis
procedures, including source detection.  Results are presented in Section~3, and
our main goal is to assess whether the {\em Chandra} sources detected are counterparts
to the {\em NuSTAR} serendips.  We also search on-line catalogs and obtain archival 
optical images to determine if the {\em Chandra} sources have optical or near-IR 
counterparts.  In Section~4, we discuss the results for the serendips with new 
{\em Chandra} coverage.

\section{Observations and Analysis}

As we are focusing on Galactic sources, these observations target {\em NuSTAR} 
serendips within $12^{\circ}$ of the Galactic plane.  Table~\ref{tab:obs} lists the 
{\em Chandra} observations that we used for this work.  We obtained seven of the 
pointings as part of our {\em Chandra} Guest Observer (GO) programs from cycle 16 
and 17, and these were carried out with the aimpoint (the center of the field of 
view) on the ACIS-S CCD chip.  We also used one additional ACIS pointing (ObsID 17704) 
from the {\em Chandra} archive.  These observations were all made during 2015 and 2016 
with exposure times ranging from 4.9 to 28.8\,ks.  As we selected the {\em Chandra}
targets before the completion of L17, the {\em Chandra} coverage only includes a
fraction of the unclassified serendips.  In the primary and secondary L17 catalogs,
there are 114 sources within $12^{\circ}$ of the Galactic plane, and 30 of them 
were classified in L17.

\begin{figure*}
\includegraphics{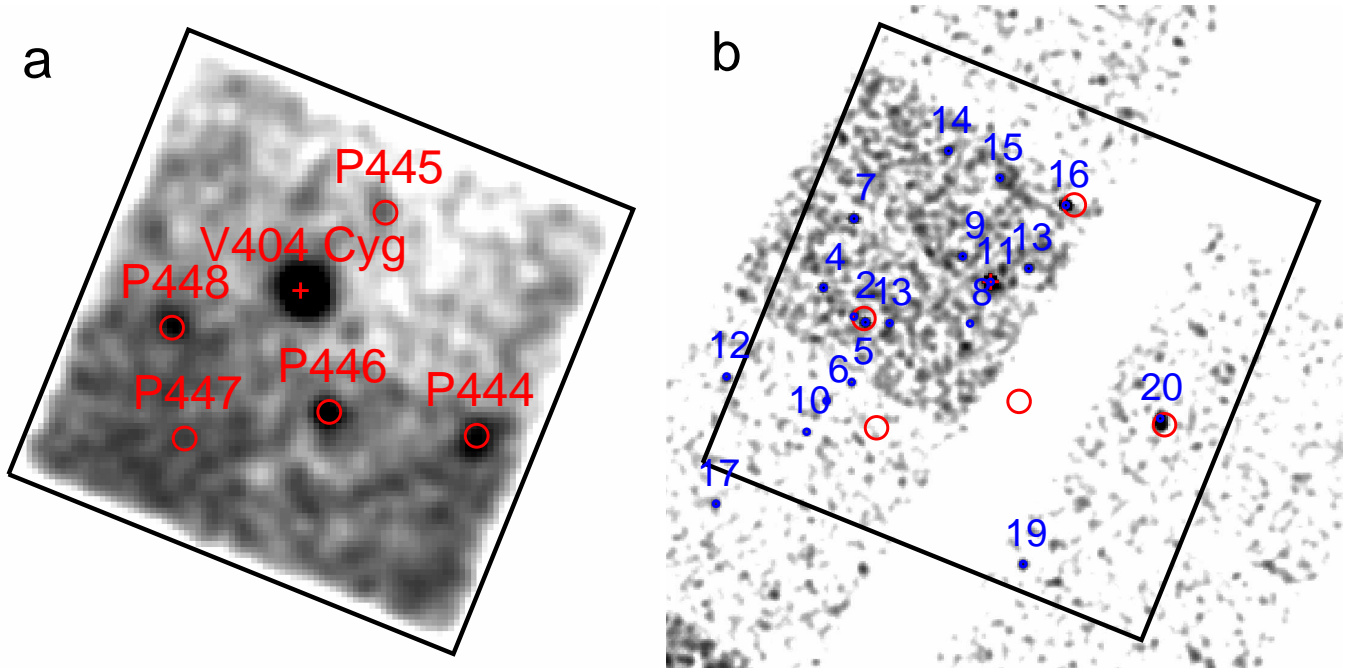}
\caption{{\em (a)} 3--24\,keV {\em NuSTAR} image for Focal Plane Module A 
from ObsID 30001010003 with an exposure time of 97\,ks.  The primary target of the observation 
was the black hole transient V404~Cyg (marked with a red ``+''), and the observation was taken 
in 2013 when V404~Cyg was in quiescence.  Five {\em NuSTAR} serendips from L17
are marked with red circles with radii of $20^{\prime\prime}$.  {\em (b)} 0.5--7\,keV {\em Chandra}
image from ObsID 17245.  The sources detected are marked with blue circles with radii of $5^{\prime\prime}$ (much larger than the actual position uncertainties).  The black square shows the {\em NuSTAR} field of view.  P446 falls in a gap between {\em Chandra} CCDs.  These images and all the images in this work are oriented so that North is up and East is to the left.\label{fig:nustar_and_chandra}}
\end{figure*}

Each of the {\em Chandra} GO observations targeted a {\em NuSTAR} serendip, and the
black hole transient V404~Cyg was the target of ObsID 17704.  In addition to the 
primary target, the {\em Chandra} observations cover other {\em NuSTAR} serendips
because of the similar sizes of the {\em Chandra} and {\em NuSTAR} fields of view, 
and Table~\ref{tab:serendips} lists the 19 serendips for which {\em Chandra} 
coverage was obtained.  Each serendip has a {\em NuSTAR} source name as well as 
a catalog number.  Sources starting with a ``P'' are in the primary L17 catalog,
and sources starting with an ``S'' are in the secondary L17 catalog.  The specific 
targets of the {\em Chandra} observations are listed in Table~\ref{tab:targets}.
Here, we have grouped the ObsIDs according to the six {\em NuSTAR} fields being 
covered.  An example field is shown in Figure~\ref{fig:nustar_and_chandra}, where 
five serendips were detected by {\em NuSTAR} (P444, P445, P446, P447, and P448).  
For {\em Chandra} ObsID 17245, the primary target was P448, and 
Figure~\ref{fig:nustar_and_chandra}b shows that four of the five serendips were 
covered.  We added ObsID 17704 (not shown in Figure~\ref{fig:nustar_and_chandra})
to this study since it provides coverage of the fifth serendip (P446).  

To analyze the data from the {\em Chandra} observations, we used the {\em Chandra} 
Interactive Analysis of Observations (CIAO) version 4.9 software and Calibration Data 
Base (CALDB) 4.7.4.  For each ObsID, we made event lists with {\ttfamily chandra\_repro} 
and searched for sources using {\ttfamily wavdetect} \citep{freeman02}.  We followed 
the source detection procedures recommended for CIAO 
users\footnote{See http://cxc.harvard.edu/ciao/threads/wavdetect/}.  Using 
{\ttfamily fluximage}, we produced a 0.5--7\,keV ``broad'' band exposure corrected
image and an exposure map for 2.3\,keV photons.  The size of the point spread function
(psf) depends strongly on the off-axis angle, and we made a psf map for an energy of
2.3\,keV and an encircled energy of 0.393.  We ran {\ttfamily wavdetect} with wavelet
scales of 1, 2, 4, 8, and 16 pixels, and set the detection threshold at a level
estimated to give the detection of one false source.  The data for ObsID 17704
was obtained when V404~Cyg was in outburst, and the dust scattering halo produces
soft X-ray emission covering much of the ACIS field of view \citep{heinz16}.
Therefore, for this ObsID, we used the ``hard'' energy band (2--7\,keV) to minimize 
the contribution from the dust scattering halo for this observation.  As described in 
the Appendix, we cross-correlated the positions of the detected {\em Chandra} sources 
with those in several OIR source catalogs (e.g., {\em Gaia}).  Where possible, we 
used the optical or infrared positions to register the {\em Chandra} images (see
Appendix for details).  However, the errors on the {\em Chandra} positions in this
work still assume the standard value of $0.64^{\prime\prime}$ (90\% confidence) for
the systematic component \citep{weisskopf05}.

For the {\em Chandra} photometry, we used {\ttfamily mkpsfmap} to determine the
95\% encircled energy radii for each source found with {\ttfamily wavdetect}
and extracted the counts in the 0.5--2\,keV, 2--7\,keV, and 0.5--7\,keV energy
bands within the circles.  We estimated the background rates by extracting counts
in the same three energy bands from large source-free regions.  As some of the
sources were on front-illuminated CCDs while others were on back illuminated
CCDs, we determined background rates for both cases, and then subtracted the
background for all sources.  All of the sources with positive numbers of
0.5--7\,keV source counts after background subtraction are included in 
Table~\ref{tab:sourcelist}.

\section{Results}

\subsection{Chandra candidates for NuSTAR serendips}

\begin{figure*}
\includegraphics{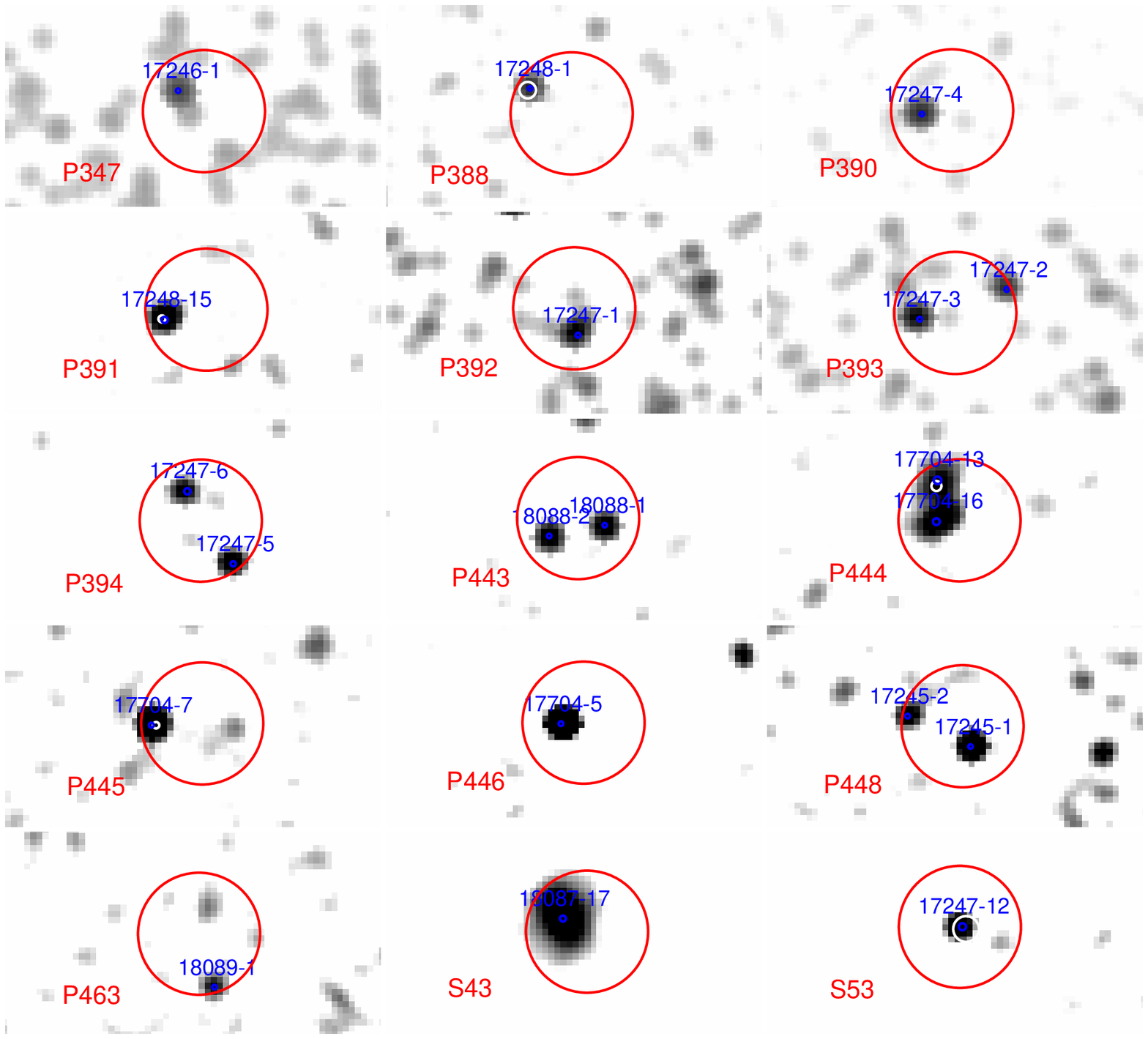}
\caption{{\em Chandra} images for the sources in Table~\ref{tab:sourcelist_candidates}.  The $20^{\prime\prime}$ {\em NuSTAR} error circles are shown in red.  The {\em Chandra} sources are marked with blue or white circles with radii equal to the position uncertainty.  The blue circles are labeled with the {\em Chandra} source names, and these also indicate which ObsIDs are used to make the images.  The white circles are for sources detected in more than one ObsID (17247-19 for P388, 17247-10 for P391, 17245-20 for P444, 17245-16 for P445, and 17248-28 for S53).  The energy band is 0.5--7\,keV except for ObsID 17704 (P444, P445, and P446) where 2--7\,keV is used.\label{fig:chandra15}}
\end{figure*}

From the full list of {\em Chandra} sources (Table~\ref{tab:sourcelist}), we consider 
sources within $20^{\prime\prime}$ of the best positions of the {\em NuSTAR} serendips 
to be potential candidates.  We use $20^{\prime\prime}$ as our criterion based on the
fact that the 90\% confidence errors on the {\em NuSTAR} positions range from
$14^{\prime\prime}$ to $22^{\prime\prime}$ depending on the significance of the
detection (L17).  Of the 19 {\em NuSTAR} serendips covered by the 
{\em Chandra} observations, there are 15 with at least one {\em Chandra} source 
within $20^{\prime\prime}$, and the {\em Chandra} sources that are candidates
for being associated with the {\em NuSTAR} serendips are listed in
Table~\ref{tab:sourcelist_candidates}.  In 10 cases, there are single candidates,
and in the other 5 cases, there are two candidates.  Figure~\ref{fig:chandra15}
shows the {\em Chandra} images, indicating that P393, P394, P443, P444, and P448
are the serendips with two candidates.

One of the serendips, S43, is the known HMXB 2RXP~J130159.6--635806 \citep{krivonos15}.
It has already been considered in the context of the Galactic populations present
in the group of {\em NuSTAR} serendips \citep{tomsick17}.  In the {\em Chandra}
program, we did not specifically target it but it was serendipitously covered
by the S44 pointing (see Table~\ref{tab:targets}).

\begin{figure*}
\includegraphics{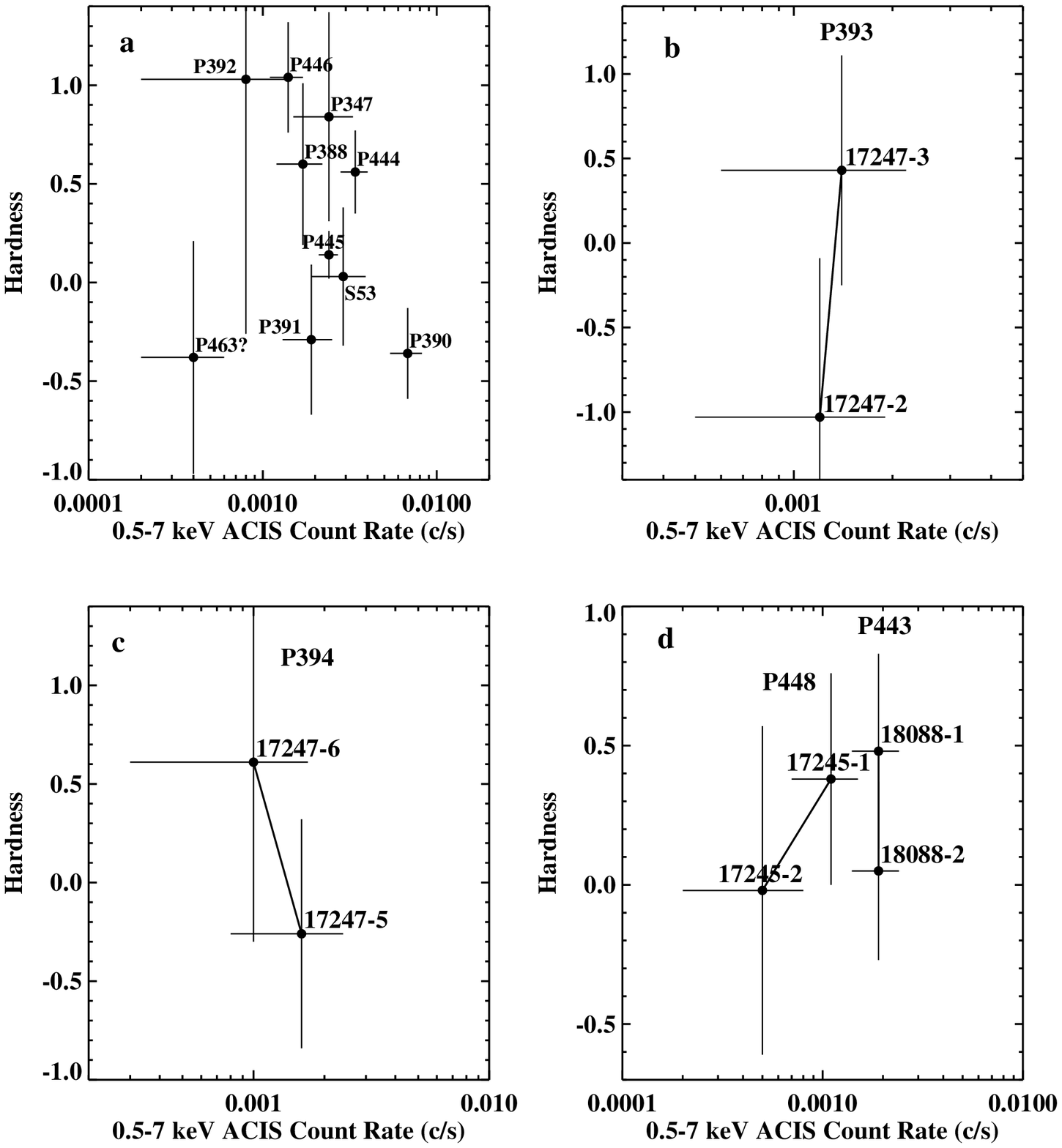}
\caption{\footnotesize Hardness-intensity diagram using 0.5--7\,keV, 0.5--2\,keV 
(=$r_{\rm soft}$), and 2--7\,keV (=$r_{\rm hard}$) {\em Chandra}/ACIS count rates.  
The hardness is defined as ($r_{\rm hard}$--$r_{\rm soft}$)/($r_{\rm hard}$+$r_{\rm soft}$).
The errors bars are 1-$\sigma$ errors using the Gehrels approximation for Poisson
statistics.\label{fig:hi}}
\end{figure*}

To determine which of the {\em Chandra} candidate counterparts are likely to be
matches to the {\em NuSTAR} serendips, we have produced hardness-intensity diagrams
(see Figure~\ref{fig:hi}).  The 0.5--7\,keV count rates are simply the counts from
Table~\ref{tab:sourcelist_candidates} divided by the exposure time for the appropriate
ObsID, and we determined the 0.5--2\,keV and 2--7\,keV rates, $r_{\rm soft}$ and
$r_{\rm hard}$, in the same manner.  The hardness is defined as the ratio of
$r_{\rm hard}$--$r_{\rm soft}$ over $r_{\rm hard}$+$r_{\rm soft}$; thus, a source with
all of its counts in the 2--7\,keV band will have a hardness of +1.0, and a source
with all of it counts in the 0.5--2\,keV band will have a hardness of $-$$1.0$.
The sources which will contribute significantly to the 3--24\,keV fluxes detected
by {\em NuSTAR} are expected to occupy the harder and/or higher count rate parts
of the plots unless they have significant variability.  For sources detected 
in multiple ObsIDs, we calculated weighted averages of the rates and the hardness 
ratios, and these are plotted in Figure~\ref{fig:hi}.

Figure~\ref{fig:hi}a shows the hardness-intensity diagrams for the serendips with 
single {\em Chandra} candidates.  Indeed, 9 of the 10 sources have either high
count rates or hard spectra.  Of the 9, P391 is the most marginal, but the
{\em Chandra} position confirms the match with the optical source that has 
previously been classified as an AGN based on its optical spectrum, providing a 
reason to think that the {\em Chandra} candidate is likely to be the correct
counterpart.  The {\em Chandra} candidate for P463 is a very soft source with 
the lowest count rate in the group, and we suspect that it (18089-1) may not
be associated with P463.  Although Figure~\ref{fig:chandra15} shows two possible 
candidates for P444 (17704-13 and 17704-16), the contamination from the V404~Cyg 
X-ray halo in ObsID 17704 does not allow for a determination of the hardness for
these two sources.  Thus, for P444, Figure~\ref{fig:hi}a shows the count rate
and hardness for 17245-20, which is a blend of 17704-13 and 17704-16.  It is 
currently unclear whether P444 is a combination of emission from two point sources 
or if there is a single extended source.  The HMXB S43 has a 0.5--7\,keV rate of 
$0.410\pm 0.007$\,c/s (outside the range of the plots) and a hardness of $0.75\pm 0.02$ 
(1-$\sigma$ errors).  

Figures~\ref{fig:hi}b, \ref{fig:hi}c, and \ref{fig:hi}d provide the values for
the serendips with two {\em Chandra} candidate counterparts.  While the {\em NuSTAR}
flux for these serendips (P393, P394, P443, and P448) may include contributions 
from both {\em Chandra} sources, the plots suggest that 17247-3, 17247-6, 
18088-1, and 17245-1 are likely to contribute more to the fluxes of their respective
{\em NuSTAR} serendips as they are the harder sources.  However, we note that the
error bars overlap in all four cases, so this is not a strong conclusion.

The four serendips without {\em Chandra} detections are P389, P395, P447, and S44.
In the first three cases, there are no {\em Chandra} sources even within an arcminute, 
indicating {\em Chandra} non-detections for P389, P395, and P447.  Explanations
for the non-detections could be source variability or possibly that some of the
{\em NuSTAR} detections are spurious.  For S44, there are two {\em Chandra} sources
that are $35^{\prime\prime}$ away from the {\em NuSTAR} position.  The angular separation
is too large to consider these as likely candidates, but we mention them as possible
candidates.

L17 used soft X-ray data, primarily from archives, to search for counterparts in
the {\em NuSTAR} error circles of the serendips.  For the 19 serendips that we are
studying in this work, {\em XMM-Newton} sources were found in 13 cases, {\em Chandra} 
sources were found in 2 cases (P347 and S53), a {\em Swift} X-ray telescope source was 
found in 1 case (P443), and there were 3 serendips with no soft X-ray counterparts 
(P389, P447, and P463).  The new {\em Chandra} positions that we are reporting here
(Table~\ref{tab:sourcelist_candidates}) are a significant improvement over {\em XMM}
and {\em Swift} because of {\em Chandra's} superior angular resolution.  For the
two serendips that already had {\em Chandra} positions, we compared the new positions
to the X-ray positions given in L17.  For P347 and S53, the differences in position 
are $0.20^{\prime\prime}$ and $0.86^{\prime\prime}$, respectively.  For P347, the difference
is considerably smaller than the position uncertainty.  For S53, the difference is
likely due to the fact that we have registered the images for the careful analysis
carried out in this work (see Appendix).

\clearpage

\subsection{Optical counterparts}

The {\em Chandra} positions allow us to check on the optical counterparts that 
were suggested in L17 based mostly on {\em XMM} positions.  In many cases, optical 
spectroscopy was performed, and the sources have been classified.  Of the 15 serendips 
listed in Table~\ref{tab:sourcelist_candidates}, eight of them have candidate OIR
counterparts in L17.  The separations between the {\em Chandra} positions 
and the L17 OIR positions are given in Table~\ref{tab:serendips_oir}.
For P347, P388, P390, and P391, the {\em Chandra} positions confirm the optical 
candidates.  In these cases, the {\em Chandra} positions are consistent with the 
optical positions with separations of $0.38^{\prime\prime}\pm 0.73^{\prime\prime}$, 
$0.29^{\prime\prime}\pm 0.74^{\prime\prime}$, $0.31^{\prime\prime}\pm 0.73^{\prime\prime}$, and 
$0.30^{\prime\prime}\pm 1.15^{\prime\prime}$ (90\% confidence errors, including statistical 
and systematic contributions) for the four sources, respectively.  All four of these 
sources have been optically classified as AGN.

\begin{figure*}
\centering
\begin{minipage}[l]{0.47\textwidth}
\includegraphics[width=\textwidth]{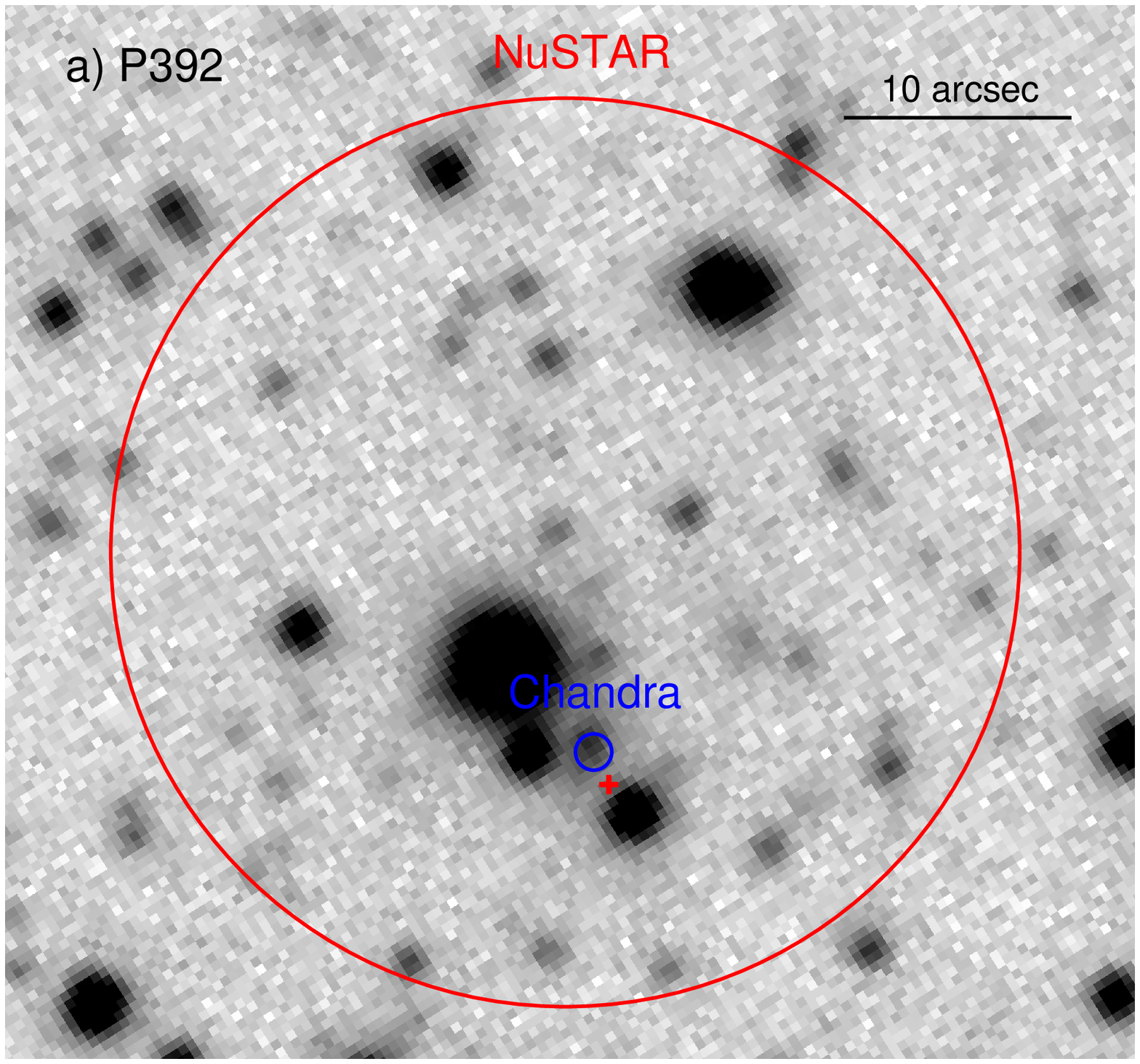}
\end{minipage}
\begin{minipage}[l]{0.45\textwidth}
\includegraphics[width=\textwidth]{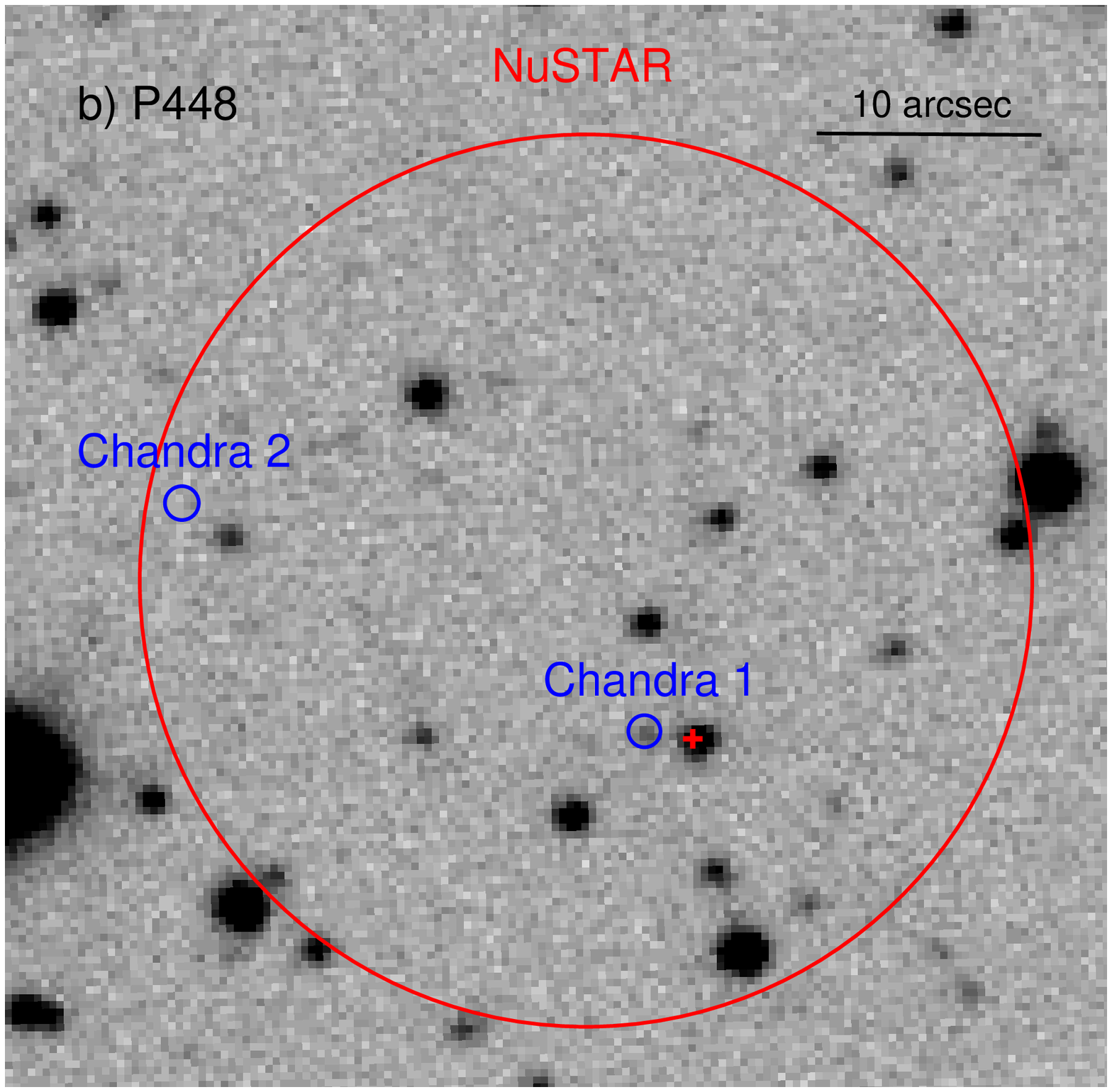}
\end{minipage}
\begin{minipage}[l]{0.47\textwidth}
\includegraphics[width=\textwidth]{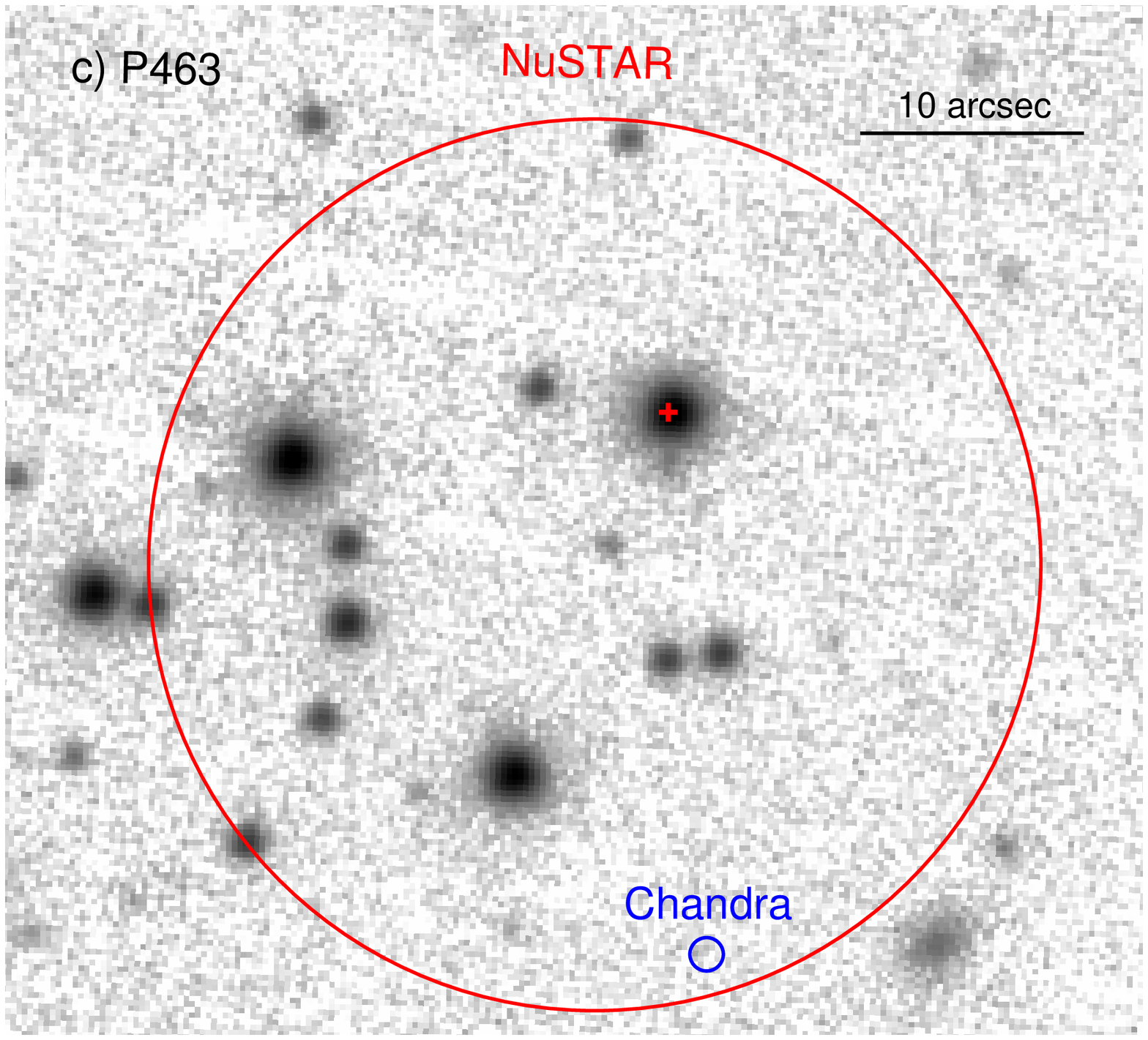}
\end{minipage}
\begin{minipage}[l]{0.45\textwidth}
\includegraphics[width=\textwidth]{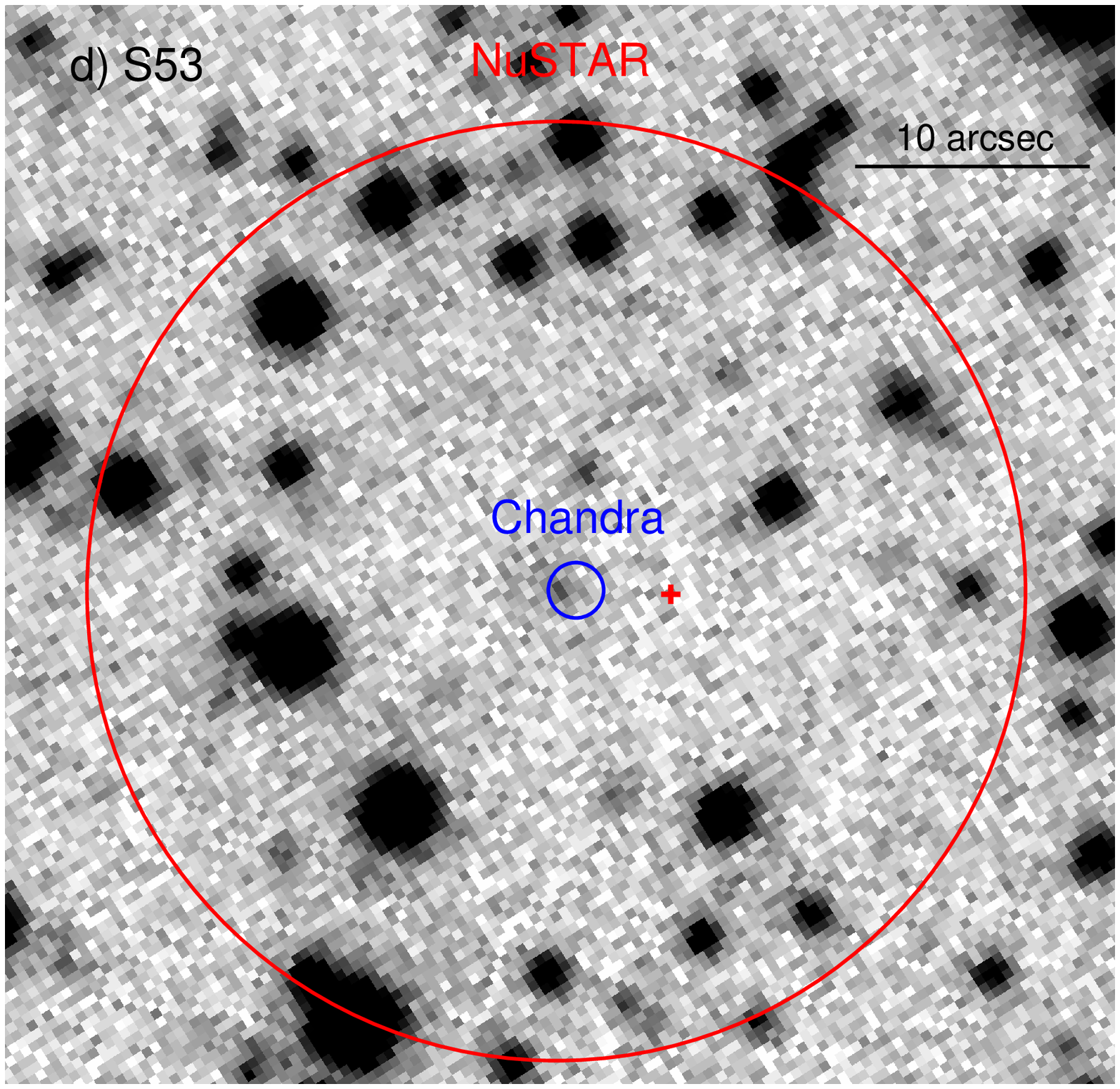}
\end{minipage}
\caption{Optical images for the cases where the {\em Chandra} positions do not
confirm the optical positions provided in L17.  The $i$-band images
come from the Sloan Digitized Sky Survey ({\em a} and {\em d}), the 
IPHAS survey ({\em b}), and PanSTARRS ({\em c}).  The red '+' symbols mark the
optical positions given in L17, and the blue circles show the
locations of X-ray sources detected by {\em Chandra}.}
\label{fig:nonconfirmations}
\end{figure*}

The {\em Chandra} positions for P392, P448, P463, and S53 are significantly
offset from the optical positions listed in L17 (see Table~\ref{tab:serendips_oir}).
Figure~\ref{fig:nonconfirmations} shows archival optical ($i$-band) images covering
the {\em NuSTAR} error regions for these four serendips.  
For P392, the {\em Chandra} source is $1.59^{\prime\prime}\pm 0.80^{\prime\prime}$ from 
the L17 position, which is based on the position of a USNO-B1.0 optical source with
$I = 15.6$ that is inside the {\em NuSTAR} error circle as well as being inside the
error circle of an {\em XMM-Newton} source.  While the VizieR data base does not have 
any optical sources with positions that are consistent with the {\em Chandra} position, 
the USNO-B1.0 source position is within $1.6^{\prime\prime}$.  The SDSS optical image 
(Figure~\ref{fig:nonconfirmations}a) suggests that the USNO-B1.0 source is actually 
a blend of optical sources, and the {\em Chandra} position shows that the correct 
P392 counterpart is the fainter source to the Northeast.  L17 obtained an optical
spectrum targeting the position of the USNO-B1.0 source and classified P392 as an
AGN with a redshift of $z = 0.197$, but there is uncertainty as to which of the
narrowly offset optical sources corresponds to the AGN.

For P448, the L17 serendip catalog indicated that there is uncertainty about the
optical counterpart.  The L17 optical position identifies a unique optical source, 
but the {\em Chandra} position indicates that a different optical source is the 
actual counterpart.  This is due to the deeper optical imaging being used here.
A search of the VizieR data base shows that the optical counterpart for P448 is
in the IPHAS catalog with a brightness of $i = 19.56\pm 0.10$, and we performed
follow-up optical spectroscopy of this source (see Section 4).  For P463, there
is no evidence for X-rays from the L17 optical source, and the only {\em Chandra}
candidate counterpart does not have an optical counterpart.  For S53, there is an
optical source consistent with the {\em Chandra} position
(Figure~\ref{fig:nonconfirmations}d).  Although S53 does not appear in any optical 
catalogs in the VizieR data base, there is a WISE source within $0.61^{\prime\prime}$ 
of the {\em Chandra} position for S53.  Specifically, WISE~J172822.82--142124.7 
has magnitudes of $m_{\rm 3.35 microns}=15.16\pm 0.04$, $m_{\rm 4.6 microns} = 14.56\pm 0.06$, 
and $m_{\rm 11.6 microns} = 11.71\pm 0.25$.  

\begin{figure*}
\begin{minipage}[l]{0.45\textwidth}
\includegraphics[width=\textwidth]{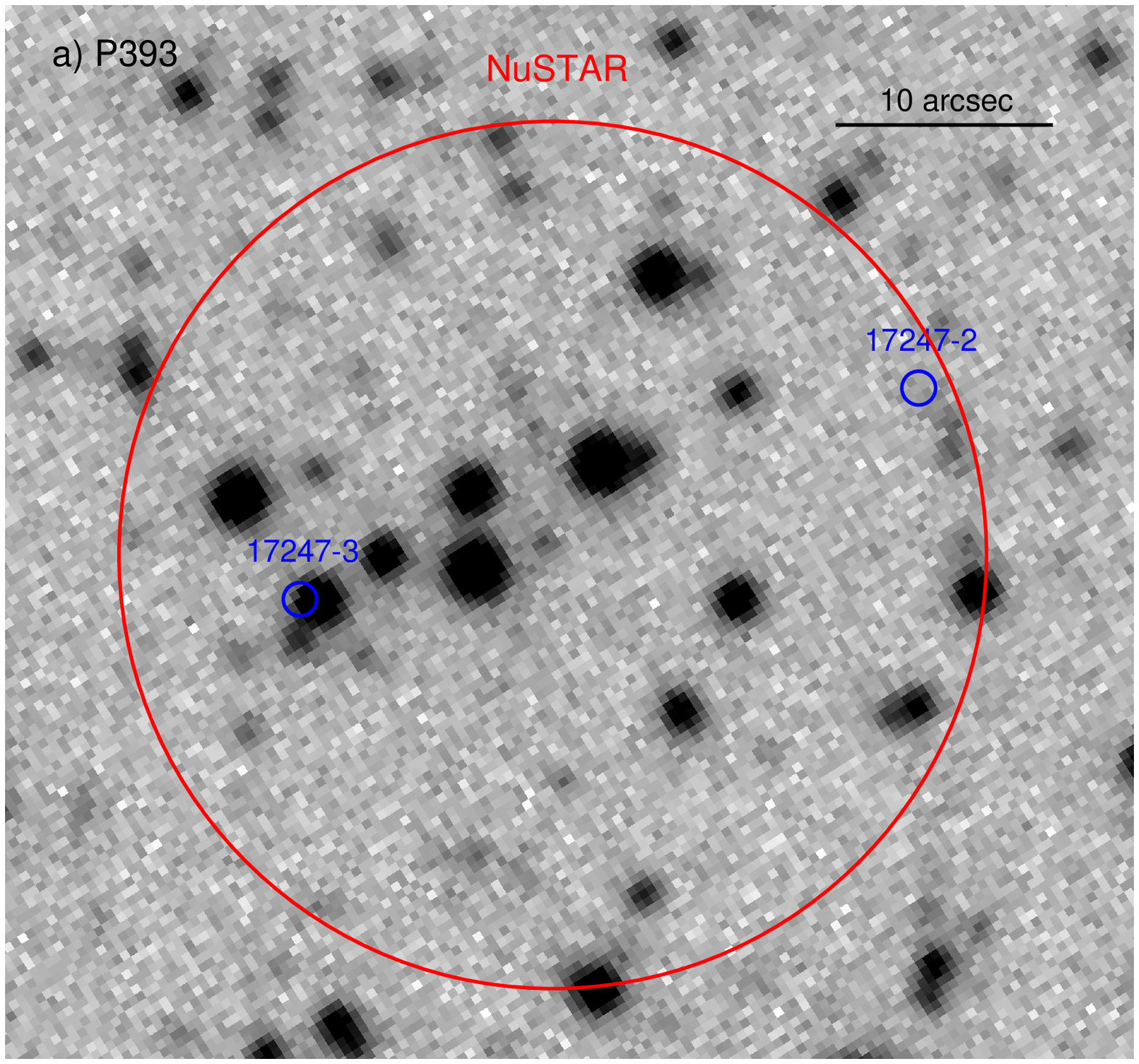}
\end{minipage}
\begin{minipage}[l]{0.48\textwidth}
\includegraphics[width=\textwidth]{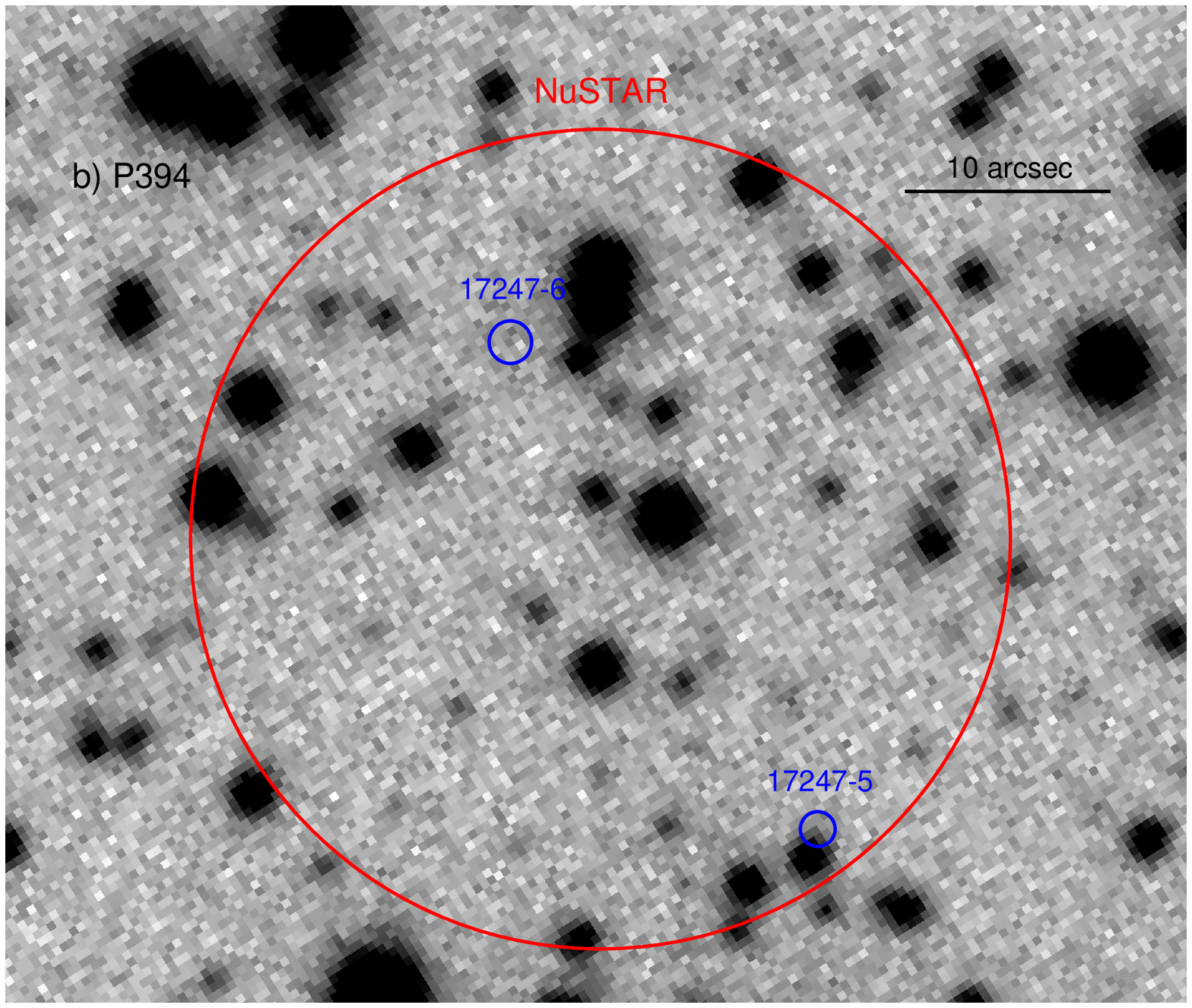}
\end{minipage}
\begin{minipage}[l]{0.45\textwidth}
\includegraphics[width=\textwidth]{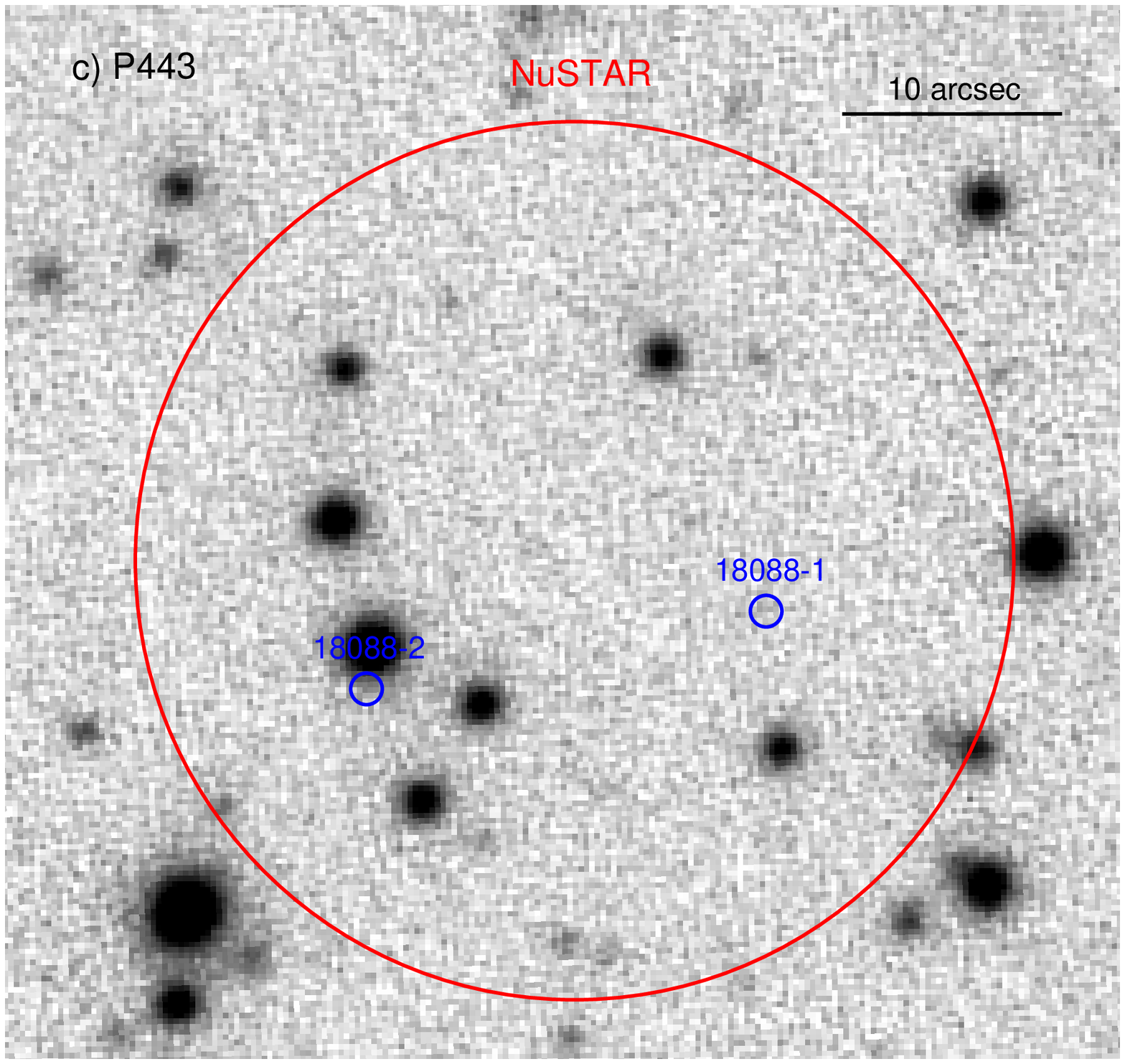}
\end{minipage}
\begin{minipage}[l]{0.47\textwidth}
\includegraphics[width=\textwidth]{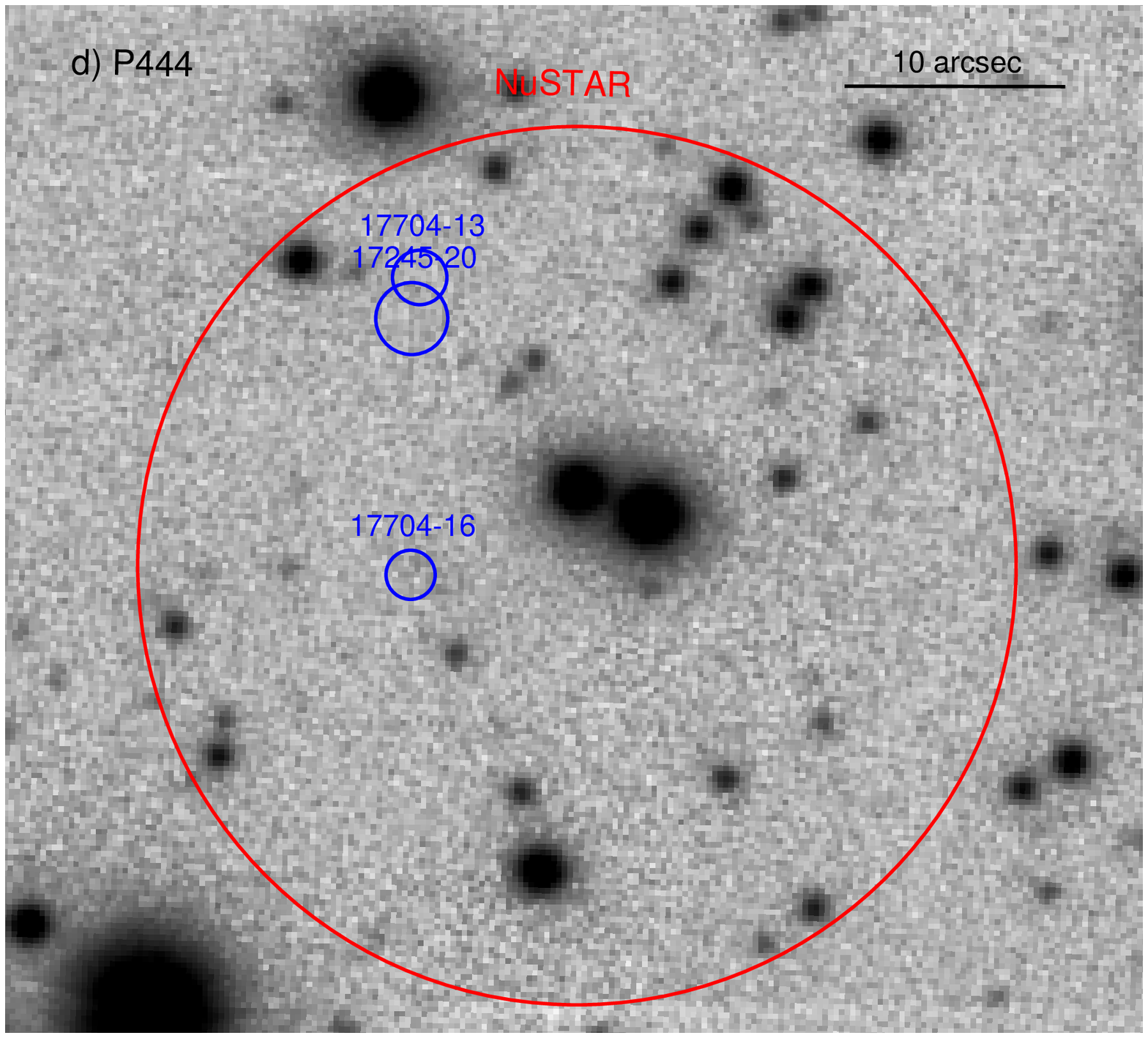}
\end{minipage}
\begin{minipage}[l]{0.45\textwidth}
\includegraphics[width=\textwidth]{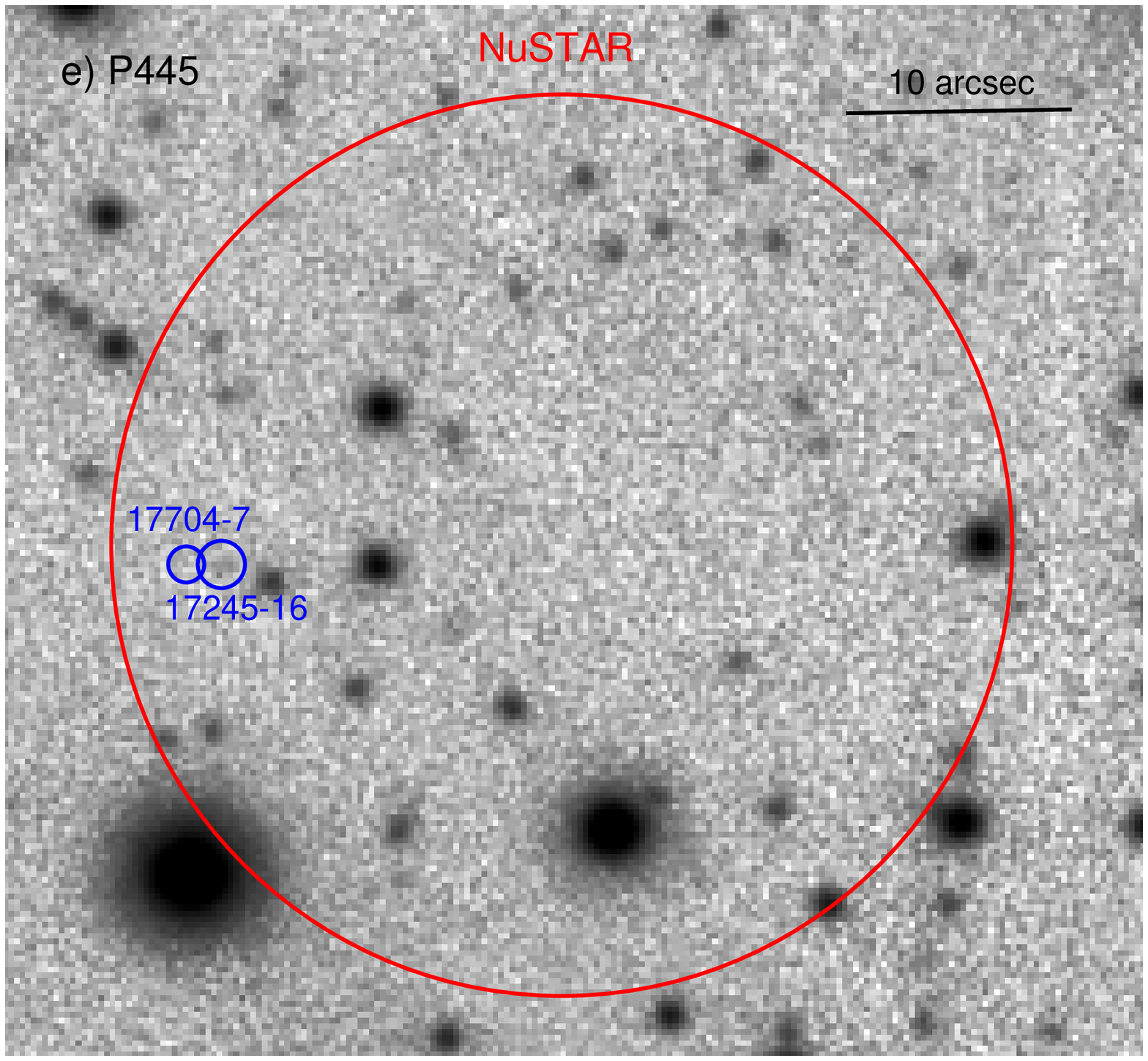}
\end{minipage}
\hspace{1.2cm}
\begin{minipage}[l]{0.46\textwidth}
\includegraphics[width=\textwidth]{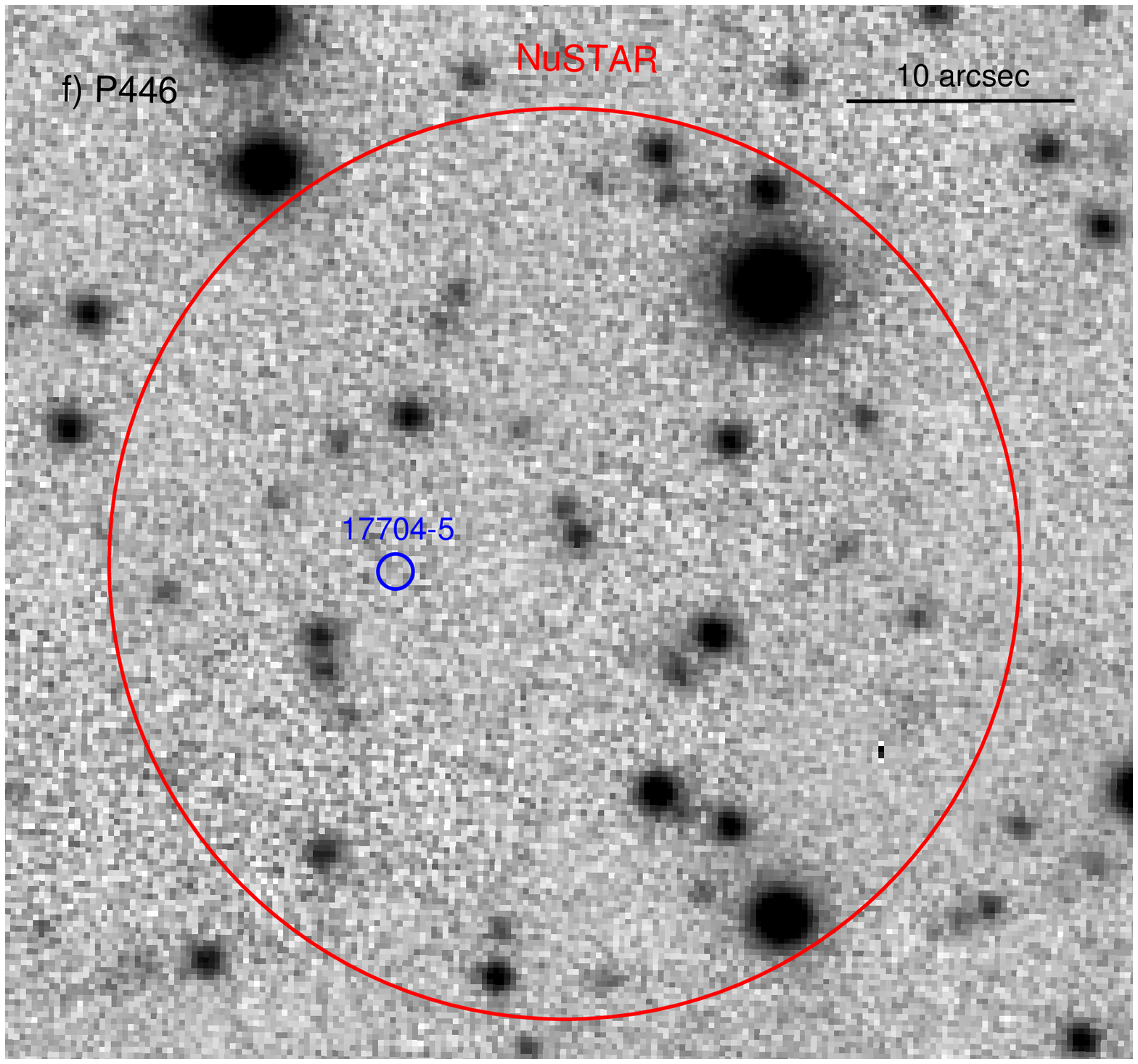}
\end{minipage}
\caption{The $i$-band images for the six sources with {\em Chandra} 
detections but no optical counterpart listed in L17.  The images come 
from the Sloan Digitized Sky Survey ({\em a} and {\em b}) and PanSTARRS 
({\em c}, {\em d}, {\em e}, and {\em f}).  The blue circles show the 
locations of X-ray sources detected by {\em Chandra}.}
\label{fig:other_optical}
\end{figure*}

Sources P393, P394, P443, P444, P445, and P446 do not have optical counterparts 
listed in L17.  We searched the catalogs in the VizieR data base for any optical 
or infrared sources consistent with the positions of the {\em Chandra} sources 
detected in the {\em NuSTAR} error circles of these serendips, and we also
show the $i$-band images for these six sources in Figure~\ref{fig:other_optical}.
With one exception, the VizieR searches did not uncover likely optical or infrared 
counterparts.  For 17247-3, which is the most likely counterpart to P393, 
2MASS~J17280709--1420245 is a near-IR source within $0.78^{\prime\prime}$ of the 
{\em Chandra} position with magnitudes of $J = 16.30\pm 0.14$ and $K_{s} = 15.47\pm 0.18$.  
The fact that the {\em Chandra} position uncertainty is $0.77^{\prime\prime}$ indicates 
that the 2MASS source is at the edge of the error circle, and this can also be seen 
in the $i$-band image from SDSS (Figure~\ref{fig:other_optical}a).  For P394, 
Figure~\ref{fig:other_optical}b shows that there is an $i$-band source relatively 
close to the {\em Chandra} source 17247-5, but its {\em Gaia} position is 
$1.5^{\prime\prime}$ from the {\em Chandra} position, making the association unlikely.  
For the other {\em Chandra} sources in the P394, P443, P444, and P446 fields, there 
are no candidate counterparts in the VizieR database or in the $i$-band images
(Figure~\ref{fig:other_optical}b, \ref{fig:other_optical}c, \ref{fig:other_optical}d,
\ref{fig:other_optical}f).  However, there may be an $i$-band counterpart for the
{\em Chandra} source associated with P445.  Figure~\ref{fig:other_optical}e
shows the PanSTARRS $i$-band image, and this is a case where the same {\em Chandra}
source was detected in two ObsIDs (17704 and 17245), but the {\em Chandra} error
circles just barely overlap.  The $i$-band source is in the middle of the 17245-16
error circle, but it is on the edge (or maybe just outside) of the 17704-7 error
circle.  In all, the $i$-band and {\em Chandra} positions may be consistent, and
we consider the $i$-band source to be a potential counterpart to P445.

Although a more detailed discussion of the results is presented in Section 5, 
here we provide a brief summary of the status of optical counterparts after using
the information learned from the {\em Chandra} positions.  In five cases, the
{\em Chandra} positions are consistent with the OIR positions reported in L17: 
P347, P388, P390, and P391 are AGN; and S43 is an HMXB.  In three cases
(P392, P448, and S53), the {\em Chandra} positions clearly identify the optical
counterpart.  Finally, in two cases (P393 and P445), there are candidate optical
counterparts.

\section{Optical Follow-Up}

\begin{figure*}
\begin{minipage}[l]{0.5\textwidth}
\includegraphics[width=\textwidth]{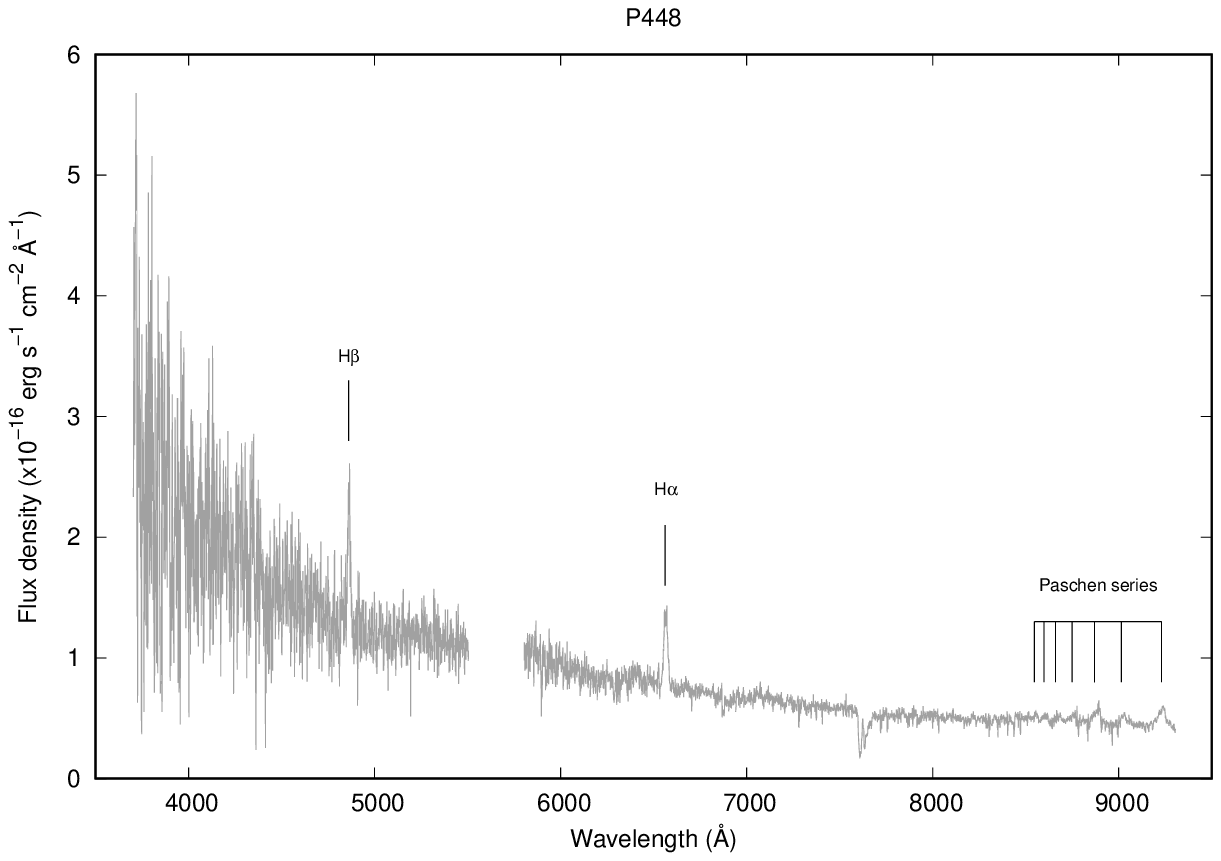}
\end{minipage}
\begin{minipage}[l]{0.5\textwidth}
\includegraphics[width=\textwidth]{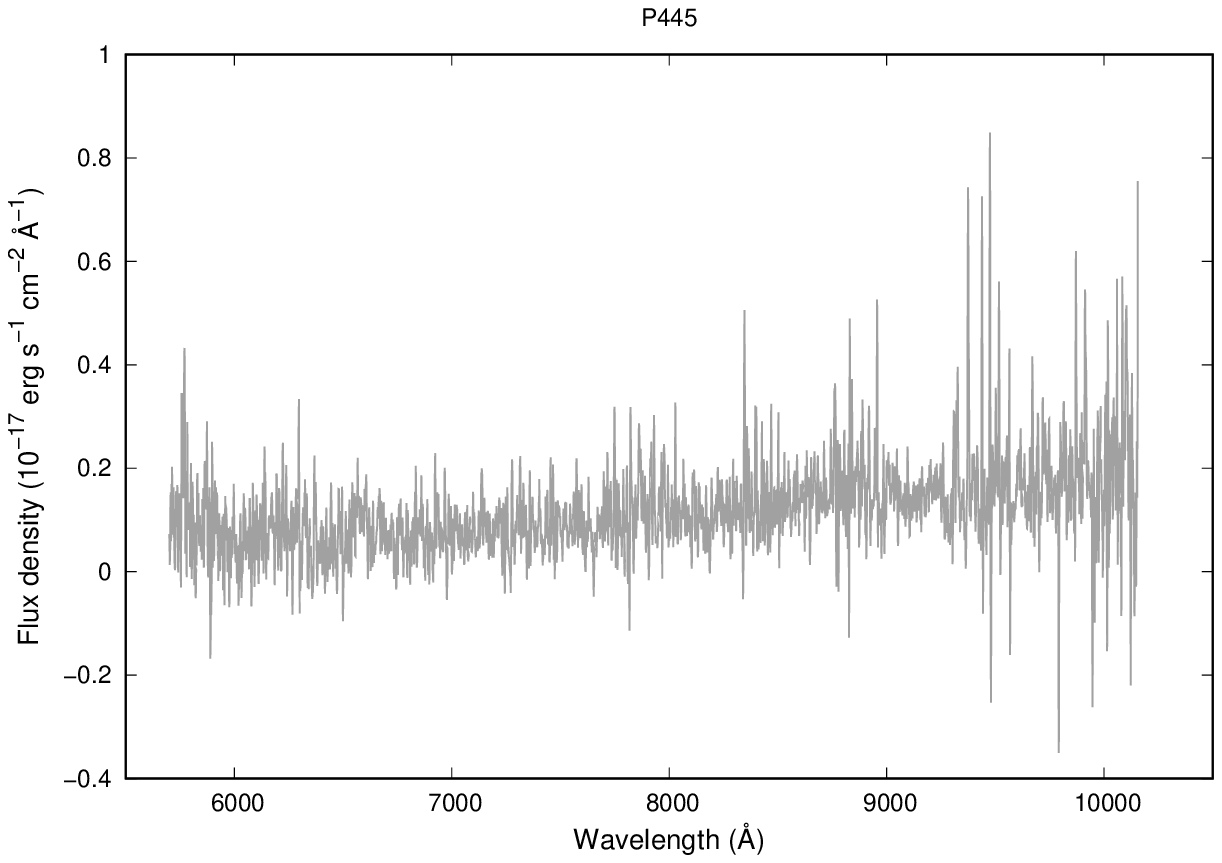}
\end{minipage}
\caption{{\em a (left)} Dereddened optical spectrum for P448 (NuSTAR~J202421+3350.9),
which we identify with the optical source IPHAS~J202421.67+335050.1. {\em b (right)} 
Optical spectrum (not dereddened) for a possible optical counterpart to P445.  The 
signal to noise for the P445 spectrum is not sufficient to detect any emission or
absorption lines.  Both spectra are from the LRIS instrument at Keck Observatory.}
\label{fig:ospec}
\end{figure*}

We obtained optical spectra for two of the sources with newly identified counterparts
(P448 and P445).  For P448, the optical counterpart to {\em Chandra} source 17245-1
is clear (see Figure~\ref{fig:nonconfirmations}b), the magnitude is $i = 19.56\pm 0.10$,
and the name of the optical source is IPHAS~J202421.67+335050.1.  We took the optical
spectrum at Keck Observatory with the Low Resolution Imaging Spectrometer (LRIS) on
2018 May 11 with an exposure time of 1800\,s.  After reducing the spectrum, we
dereddened it based on the Galactic extinction along the line of sight to the source
of $N_{\rm H} = 6.6\times 10^{21}$\,cm$^{-2}$ \citep{kalberla05}, which corresponds to
$A_{V} = 3.0$ \citep{go09}, and the spectrum is shown in Figure~\ref{fig:ospec}a.  We
detect H$\alpha$, H$\beta$, and Paschen series emission lines at zero redshift,
indicating that P448 is a Galactic source.  Table~\ref{tab:lines} provides the 
detailed line properties for H$\alpha$ and H$\beta$ taken from the optical spectrum 
before dereddening.  We discuss the nature of the source, considering the optical
and X-ray properties in Section 5.2.

We also used Keck/LRIS to observe a very faint
$i$-band source that may be the counterpart to P445.  The observation occurred on
2017 September 16 with an exposure time of 1800\,s.  As described above, the optical
source is just outside the {\em Chandra} error circle for 17704-7 (see 
Figure~\ref{fig:other_optical}e), but it is coincident with 17245-16.  The LRIS
spectrum is shown in Figure~\ref{fig:ospec}b.  While we detect a somewhat reddened
continuum, the signal-to-noise is too poor to conclude on whether there are any
emission or absorption lines, and it is not possible to classify the source or even
to say whether it is Galactic or extragalactic.

\section{Discussion}

Here, we discuss the results that we have obtained for 19 {\em NuSTAR} serendips
within $12^{\circ}$ of the Galactic plane, using {\em Chandra} observations, archival
optical observations, optical and near-IR (OIR) catalog information, and follow-up
optical spectroscopy.  We discuss the sources in the following groupings:
1. confirmations of previous classifications; 2. sources with {\em Chandra}
detections and a possible OIR counterpart; 3. sources with {\em Chandra} detections
but no detected OIR counterpart; and 4. sources without {\em Chandra} detections.

\subsection{Group 1: Confirmations (P347, P388, P390, P391, and S43)}

The first four of these sources are AGN, and S43 is the HMXB 2RXP~J130159.6--635806
\citep{krivonos15}.  The {\em Chandra} positions confirm the classifications.
For the AGN, the {\em Chandra} sources with the best positions are: 17246-1, 17248-1, 
17247-4, and 17248-15, respectively (see Table~\ref{tab:sourcelist_candidates}).  
With these confirmations, the classification work on these sources should be complete.

\subsection{Group 2: Chandra and OIR (P392, P393, P445, P448, and S53)}

For P392, L17 show an optical spectrum that is clearly an AGN with $z = 0.197$;
however, the optical coordinates given by L17 are offset by
$1.59^{\prime\prime}\pm 0.80^{\prime\prime}$ from the {\em Chandra} position 
for 17247-1.  Given the fact that the P392 field is crowded, and we cannot confirm 
that the correct counterpart was targeted, it would be advisable to obtain another
optical spectrum to confirm the AGN classification.

For P393, the most likely {\em Chandra} counterpart is 17247-3, and a possible OIR
identification is 2MASS~J17280709--1420245.  Although the {\em Chandra} and 2MASS
sources are separated by $0.78^{\prime\prime}\pm 0.77^{\prime\prime}$, and the identification
is not certain, we suggest that it would be worthwhile to obtain an optical or near-IR
spectrum of the 2MASS source before obtaining a deeper image to look for other potential
counterparts.  The 2MASS magnitudes are $J = 16.30\pm 0.14$ and $K_{s} = 15.47\pm 0.18$,
so obtaining a spectrum would not require a major effort.

We have a candidate optical counterpart for P445, and we obtained an optical spectrum.
The optical spectrum has a low signal-to-noise, and we were not able to classify the source.
Two possible next steps would be to obtain a higher quality optical spectrum or to obtain
a deeper near-IR image to search for other possible counterparts.

For P448, we were successful in identifying a unique optical counterpart and obtaining an
optical spectrum showing that this serendip is Galactic with optical emission lines.
P448 is among the group of serendips that are high priority for obtaining classifications 
to constrain the population of faint HMXBs \citep{tomsick17}.  P448 is detected in the 
8--24\,keV bandpass (L17), and, at $b$ = --2.1$^{\circ}$, it is within $5^{\circ}$ of the 
Galactic plane.  The presence of H$\alpha$ and H$\beta$ in emission would be expected
if P448 was an HMXB (e.g., a Be X-ray binary), and we have considered this possibility.
If the source is a binary with a B-type companion, it would be brighter than 
$M_{V}$ = --0.25 and $M_{I}$ = --0.13 \citep{cox00}.  For the extinction used in 
Section 4 ($A_{V} = 3.0$), $A_{I} = 1.44$, and the observed magnitude ($i = 19.56$)
implies a distance of 45\,kpc.  A much higher extinction can be ruled out by the 
optical spectrum.  Thus, an HMXB can also be ruled out because such a large distance
would put the source outside of the Galaxy.  

Although an HMXB is ruled out, with the optical emission lines, possibilities for the
nature of P448 include an LMXB or a CV.  The {\em NuSTAR} spectrum is consistent with 
being a relatively hard power-law with a photon index of $\Gamma = 1.7^{+0.4}_{-0.3}$, 
which does not clearly distinguish between the two possibilities.  The H$\alpha$
FWHM of $1057\pm 14$\,km/s (see Table~\ref{tab:lines}) also does not provide a 
clear distinction, but it does favor a CV nature because LMXBs typically have 
broader lines \citep{casares15}.  If P448 is a CV, one would expect it to have 
an absolute magnitude in the range $M_{V} = 4-11$ \citep{patterson98}.  If we assume
$M_{I}\sim 7$ and $A_{I} = 1.44$, we derive a distance of $\sim$2\,kpc.  Given the
{\em NuSTAR} flux of $9\times 10^{-14}$ erg\,cm$^{-2}$\,s$^{-1}$ (3--24\,keV), we
calculate a luminosity of $4\times 10^{31}$\,erg\,s$^{-1}$, which is also consistent
with the source being a CV.

L17 show an optical spectrum for S53, which is an AGN with $z = 0.688$.  Although 
the optical coordinates given in L17 do not match 17247-12, we found an error 
in the optical coordinates, in this exceptional case.  The coordinates given were 
the Magellan telescope pointing coordinates.  We have checked that the position of the
slit used to take the optical spectrum matches the {\em Chandra} position, and we can
consider the AGN classification of S53 to be correct.

\subsection{Group 3: Chandra but No OIR (P394, P443, P444, P446, and P463)}

P394 and P443 both have two possible {\em Chandra} counterparts.  For P394, 17247-6
may have a harder spectrum, but the errors are large.  Also, 17247-5 is slightly
brighter.  The true {\em NuSTAR} serendip may be either of these sources or possibly
a combination of the two.  Neither of these sources have optical counterparts, and
deeper near-IR observations are required for identifications.

P444 may also have two {\em Chandra} counterparts; however, they are not clearly
resolved, and P444 may possibly be a single extended source.  Given the hardness
of the source, a pulsar wind nebula (PWN) is a possibility.  The fact that the source
or sources do not have optical counterparts is consistent with a PWN nature, but this
does not constitute proof.  P444 is $7^{\prime}$ off-axis in {\em Chandra} ObsID 17704
and $9^{\prime}$ off-axis in ObsID 17245; thus, the smearing of the PSF complicates
an assessment of whether the source is extended or not.  A dedicated {\em Chandra}
observation with the source on-axis may be the next step toward classifying P444.

P446 has a unique {\em Chandra} counterpart, 17704-5, which has a hard spectrum
(all $\sim$40 ACIS counts in the 2--7\,keV band) and is almost certainly the correct
counterpart to the {\em NuSTAR} serendip.  It would be of great interest to obtain
a deeper near-IR image to look for a counterpart.

We mention that P444 may be a PWN because it may be an extended X-ray source, but it
is worth mentioning that all sources with X-ray detections but no OIR counterparts
should be considered to be pulsar or magnetar candidates.  If deeper near-IR images
do not uncover counterparts for P394, P443, P444, and P446, then the pulsar or
magnetar possibility would be likely.

Although P463 also had a unique {\em Chandra} counterpart, 18089-1 has a low X-ray
flux and a very soft X-ray spectrum.  Thus, we think there is a good chance that
18089-1 is not the correct counterpart.  There are a couple possible implications:
one is that the {\em NuSTAR} serendip may be variable, which could suggest that it
is a Galactic source.  The other possibility is that P463 is a spurious detection;
however, this is somewhat unlikely since L17 quote a false alarm probability (FAP)
in the 3--24\,keV band of $2\times 10^{-9}$ for this source.  L17 also quote a
3--24\,keV flux of $(1.54\pm 0.42)\times 10^{-13}$\,erg\,cm$^{-2}$\,s$^{-1}$ for P463.
Assuming an absorbed power-law spectral shape with a column density of
$N_{\rm H} = 10^{22}$\,cm$^{-2}$ and a photon index of $\Gamma = 2$, the L17 flux
would be expected to produce a {\em Chandra} source with 167 counts.  Such
a bright source would have definitely been strongly detected in ObsID 18089.
Comparing 167 counts to the detectability limit suggests that P463 is variable
by a factor of $\sim$20.

\subsection{Group 4: No Chandra Detection (P447, P389, P395, and S44)}

The {\em Chandra} non-detections for these four sources leave open the same possibilities
just discussed for P463: that the sources are variable or spurious.  In these cases,
the spurious possibility is more likely than for P463 because their FAPs are lower.
For P447, P389, and P395, the FAPs are $4\times 10^{-7}$, $5\times 10^{-5}$, and
$3\times 10^{-6}$, respectively, and these source detections have the lowest significances
of the sources in each field.  While L17 did not calculate FAPs for the
secondary source catalog, the flux for S44 indicates a significance of 3.3-$\sigma$.

Keeping in mind that the sources may be spurious, we use the {\em NuSTAR} fluxes to
estimate the expected number of {\em Chandra} counts, making the same assumptions in
the calculation for P463.  The 3--24\,keV flux for P447 in L17 is
$(3.7\pm 1.2)\times 10^{-14}$\,erg\,cm$^{-2}$\,s$^{-1}$.  For P389, P395, and S44, the
3--8\,keV fluxes are $(2.1\pm 0.6)\times 10^{-14}$, $(1.8\pm 0.7)\times 10^{-14}$,
and $(4.3\pm 1.2)\times 10^{-14}$\,erg\,cm$^{-2}$\,s$^{-1}$, respectively.  Among all
the {\em Chandra} ObsIDs
where these sources are covered, the predicted number of counts ranges from 11 to 58.
Thus, at the flux levels measured by {\em NuSTAR}, these sources should have been
detected in the {\em Chandra} observations unless they are variable (or spurious).

\subsection{Status of Galactic Hard X-ray Population Studies}

In \cite{tomsick17}, we developed a framework for constraining the faint HMXB
population, using the {\em NuSTAR} coverage within $5^{\circ}$ of the Galactic Plane,
and focusing on the sources detected in the 8--24\,keV band. L17 report
8--24\,keV detections for 30 serendips at $|b|$$<$$5^{\circ}$, and these sources are
our highest priority for classifications.  Six of the 30 serendips were classified 
(as 3 AGN, an HMXB, a magnetar, and an HMXB candidate) previously, and in this work, 
we have taken another step toward improving the completeness of classifications
by classifying P448 as a likely CV and by taking the next steps toward classifying
P444, P445, and P446, which are also on the high priority list.  Even if the CV
classification for P448 is not completely certain, we determined that an HMXB 
is ruled out since an HMXB would need to be at a distance of at least 45\,kpc.
Although we are still far from having a complete set of classifications for 
the {\em NuSTAR} serendipitous survery sources, when we do achieve a high 
level of completeness, we will have significant constraints on the faint populations 
of HMXBs as well as other populations that are found almost exclusively at low 
Galactic latitude, such as magnetars.

\acknowledgments

This work made use of data from the {\it NuSTAR} mission, a project led by the
California Institute of Technology, managed by the Jet Propulsion Laboratory,
and funded by the National Aeronautics and Space Administration. We thank the
{\it NuSTAR} Operations, Software and  Calibration teams for support with the
execution and analysis of these observations.  This research has made use of
the {\it NuSTAR}  Data Analysis Software (NuSTARDAS) jointly developed by the
ASI Science Data Center (ASDC, Italy) and the California Institute of Technology
(USA).  Data from {\em Chandra} were also used, and the work on serendipitous 
{\em NuSTAR} sources is partially funded by {\em Chandra} grants GO5-16154X and 
GO6-17135X.  This work has made use of data from the European Space Agency (ESA) 
mission {\em Gaia}, processed by the {\em Gaia} Data Processing and Analysis 
Consortium (DPAC).  Funding for the DPAC has been provided by national 
institutions, in particular the institutions participating in the {\em Gaia} 
Multilateral Agreement.  This research had made use of the SIMBAD database and 
the VizieR catalog access tool, CDS, Strasbourg, France as well as the VISTA 
and WISE databases and publicly available images from PanSTARRS, SDSS, and IPHAS.

\appendix

For the eight {\em Chandra} ObsIDs listed in Table~\ref{tab:obs}, we used
{\ttfamily wavdetect} for source detection as described in Section~2.  
Table~\ref{tab:sourcelist} lists all the sources that were detected with the
criterion that the number of ACIS counts minus the error on the number of
ACIS counts is a positive number.  The number of sources detected ranged from
14 for ObsID 17246 to 47 for ObsID 18089.  Table~\ref{tab:sourcelist} gives
the source number, the angular distance between the {\em Chandra} aimpoint 
and the source ($\theta$), the {\em Chandra} position and uncertainty, the 
number of ACIS counts for each source, and notes about other identifications.

We searched several on-line catalogs (e.g., {\em Gaia}, {\em WISE}, and {\em VISTA})
for matches with the {\em Chandra} sources.  If more than a few matches can be
confidently found, then the {\em Chandra} images can be registered.  We found
between 4 and 9 {\em Chandra}/OIR matches for ObsIDs 17247, 17248, 18088, 18089,
and 18087, and performed the shifts given in Table~\ref{tab:shifts}.  The largest
shift was $0.59^{\prime\prime}$ for ObsID 17247.  For ObsID 17245, the only match we
found was with V404~Cyg.  We checked that the {\em Chandra} position matched the
known position of V404~Cyg, but we did not perform any shifts.  For ObsID 17704,
we did not find OIR matches, but we did have four matches with the same X-ray
sources found in ObsID 17245, and we shifted the positions for ObsID 17704 based
on the matches between these two ObsIDs.  Although these shifts are expected to
provide a decrease in the systematic position uncertainty, we were conservative
and continued to use the standard value of $0.64^{\prime\prime}$ (90\% confidence)
for errors quoted in Table~\ref{tab:sourcelist} and elsewhere in this work.




\begin{table*}
\caption{{\em Chandra} Observations\label{tab:obs}}
\begin{minipage}{\linewidth}
\begin{center}
\begin{tabular}{ccccc} \hline \hline
ObsID & $l$ (deg) & $b$ (deg) & Start Time (UT) & Exposure time (s)\\ \hline
17245 & 73.14     & --2.15    & 2015 November 27, 3.26 h  & 13,946\\
17246 & 321.62    & +6.76     & 2015 June 12, 16.04 h     & 4,939\\
17247 & 10.29     & +11.17    & 2015 February 24, 16.15 h & 4,952\\
17248 & 10.35     & +11.27    & 2015 February 14, 9.56 h  & 5,954\\
17704 & 73.12     & --2.10    & 2015 July 25, 18.68 h     & 28,755\\
18087 & 304.25    & --0.96    & 2016 May 3, 10.42 h       & 9,956\\
18088 & 62.11     & --9.39    & 2016 May 2, 16.42 h       & 9,943\\
18089 & 80.27     & --11.18   & 2016 April 7, 7.17 h      & 19,929\\ \hline
\end{tabular}
\end{center}
\end{minipage}
\end{table*}

\begin{table*}
\caption{{\em NuSTAR} Serendips Studied in this Work\label{tab:serendips}}
\begin{minipage}{\linewidth}
\begin{center}
\begin{tabular}{cccc} \hline \hline
{\em NuSTAR} Name &  L17 Catalog & R.A. (J2000) & Decl. (J2000)\\
                  &  Name        & (degrees)    & (degrees)\\ \hline
NuSTAR J145439--5135.3 & P347 & 223.66539 & --51.58938\\
NuSTAR J172750--1414.8 & P388 & 261.96027 & --14.24671\\
NuSTAR J172755--1417.4 & P389 & 261.98068 & --14.29110\\
NuSTAR J172803--1423.0 & P390 & 262.01660 & --14.38466\\
NuSTAR J172805--1416.5 & P391 & 262.02155 & --14.27610\\
NuSTAR J172805--1420.9 & P392 & 262.02435 & --14.34984\\
NuSTAR J172806--1420.3 & P393 & 262.02646 & --14.33960\\
NuSTAR J172807--1418.2 & P394 & 262.03070 & --14.30409\\
NuSTAR J172843--1419.0 & P395 & 262.17938 & --14.31705\\
NuSTAR J202313+2042.8  & P443 & 305.80463 &  +20.71419\\
NuSTAR J202339+3347.7  & P444 & 305.91287 &  +33.79657\\
NuSTAR J202351+3354.3  & P445 & 305.96622 &  +33.90517\\
NuSTAR J202359+3348.4  & P446 & 305.99915 &  +33.80822\\
NuSTAR J202420+3347.7  & P447 & 306.08377 &  +33.79522 \\
NuSTAR J202421+3350.9  & P448 & 306.09122 &  +33.84916\\
NuSTAR J211935+3337.0  & P463 & 319.89969 &  +33.61708\\
NuSTAR J130157--6358.1 & S43  & 195.48941 & --63.96994\\
NuSTAR J130324--6348.6 & S44  & 195.85243 & --63.81157\\
NuSTAR J172822--1421.4 & S53  & 262.09515 & --14.35694\\ \hline
\end{tabular}
\end{center}
\end{minipage}
\end{table*}

\begin{table*}
\caption{{\em Chandra} Coverage of {\em NuSTAR} Serendips\label{tab:targets}}
\begin{minipage}{\linewidth}
\begin{center}
\begin{tabular}{cccc} \hline \hline
      & Primary {\em Chandra} & Target of {\em NuSTAR}  & Other Serendips\\
ObsID & Target                & Field                   & Covered\\ \hline
17245 & P448   & V404 Cyg         & P444, P445, P447\\
17704 & V404 Cyg & V404 Cyg           & P444, P445, P446, P447\\ \hline
17246 & P347  & IGR J14552--5133 & --\\ \hline
17247 & P392  & PDS 456          & P388-P391, P393-P395, S53\\
17248 & P388  & PDS 456          & P389, P391, P395, S53\\ \hline
18087 & S44   & PSR B1259--63    & S43\\ \hline
18088 & P443   & Nova Del 2013    & --\\ \hline
18089 & P463  & 2MASX J21192912+3332566 & --\\ \hline 
\end{tabular}
\end{center}
\end{minipage}
\end{table*}

\begin{table*}
\caption{{\em Chandra} Candidate Matches to {\em NuSTAR} Serendips\label{tab:sourcelist_candidates}}
\begin{minipage}{\linewidth}
\begin{center}
\begin{tabular}{lcccccc} \hline \hline
  &  & {\em Chandra} R.A. & {\em Chandra} Decl. & Position & ACIS & Serendip\\
Source ID\footnote{The sources are identified by ObsID and the {\em Chandra} source number from the full source list (Table~\ref{tab:sourcelist}).}     & CXOU Name\footnote{The naming convention for unregistered {\em Chandra} sources is for their names to start with CXOU (see http://cxc.cfa.harvard.edu/cdo/scipubs.html).} & (J2000, h, m, s)   & (J2000, deg, $^{\prime}$, $^{\prime\prime}$) & Error ($^{\prime\prime}$)\footnote{The 90\% confidence uncertainty on the position, including statistical and systematic contributions.} & Counts\footnote{The number of ACIS counts detected (after background subtraction) in the 0.5--7 keV band (except for ObsID 17704, for which the energy band is 2--7\,keV).  The errors are 68\% confidence Poisson errors using the analytical approximations from \cite{gehrels86}.} & Number\footnote{If the {\em Chandra} source falls within the error circle of a {\em NuSTAR} serendipitous source, then this indicates the number of the {\em NuSTAR} serendipitous source in the L17 catalog.}\\ \hline
17246-1 & J145440.6--513515 & 14 54 40.60 & --51 35 15.1 &  0.73 & $  11.9 \pm    4.6$ & P347\\
\hline
17248-1 & J172751.4--141440 & 17 27 51.40 & --14 14 40.0 &  0.74 & $   9.8 \pm    4.3$ & P388\\
17247-19 &  --              & 17 27 51.45 & --14 14 40.6 &  2.72 & $   8.2 \pm    4.1$ & P388\\
\hline
17247-4 & J172804.6--142306 & 17 28 04.67 & --14 23 06.0 &  0.73 & $  33.8 \pm    6.9$ & P390\\
\hline
17248-15 & J172806.1--141637 & 17 28 06.12 & --14 16 37.3 &  1.03 & $  12.4 \pm    4.7$ & P391\\
17247-10 & --                & 17 28 06.17 & --14 16 37.0 &  1.15 & $   8.8 \pm    4.1$ & P391\\
\hline
17247-1 & J172805.7--142108 & 17 28 05.76 & --14 21 08.2 &  0.80 & $   4.8 \pm    3.4$ & P392\\
\hline
17247-2 & J172805.1--142014 & 17 28 05.19 & --14 20 14.9 &  0.78 & $   5.8 \pm    3.6$ & P393\\
17247-3 & J172807.1--142024 & 17 28 07.15 & --14 20 24.6 &  0.77 & $   6.8 \pm    3.8$ & P393\\
\hline
17247-5 & J172806.6--141828 & 17 28 06.64 & --14 18 28.9 &  0.85 & $   7.9 \pm    4.0$ & P394\\
17247-6 & J172807.6--141805 & 17 28 07.67 & --14 18 05.1 &  1.04 & $   4.9 \pm    3.4$ & P394\\
\hline
18088-1 & J202312.4+204248 & 20 23 12.49 & +20 42 48.8 &  0.72 & $  18.6 \pm    5.4$ & P443\\
18088-2 & J202313.7+204245 & 20 23 13.78 & +20 42 45.3 &  0.72 & $  19.5 \pm    5.6$ & P443\\
\hline
17704-13 & J202339.6+334800 & 20 23 39.67 & +33 48 00.8 &  1.24 & $  30.0 \pm    6.9$ & P444\\
17704-16 & J202339.6+334747 & 20 23 39.70 & +33 47 47.2 &  1.12 & $  42.8 \pm    7.9$ & P444\\
17245-20 & --               & 20 23 39.69 & +33 47 58.9 &  1.64 & $  48.1 \pm    8.4$ & P444\\
\hline
17704-7 & J202353.2+335418  & 20 23 53.23 & +33 54 18.1 &  0.80 & $  35.0 \pm    7.1$ & P445\\
17245-16 & --               & 20 23 53.11 & +33 54 18.0 &  1.05 & $  50.5 \pm    8.5$ & P445\\
\hline
17704-5 & J202400.3+334829 & 20 24 00.39 & +33 48 29.3 &  0.77 & $  41.4 \pm    7.6$ & P446\\
\hline
17245-1 & J202421.6+335050 & 20 24 21.69 & +33 50 50.3 &  0.72 & $  15.4 \pm    5.1$ & P448\\
17245-2 & J202423.3+335100 & 20 24 23.33 & +33 51 00.3 &  0.76 & $   7.5 \pm    4.0$ & P448\\
\hline
18089-1 & J211935.5+333644 & 21 19 35.52 & +33 36 44.0 &  0.75 & $   8.2 \pm    4.1$ & P463\\
\hline
18087-17 & J130158.6--635807 & 13 01 58.66 & --63 58 07.5 &  0.97 & $4141.4 \pm   65.6$ & S43\\ 
\hline
17247-12 & J172822.7--142124 & 17 28 22.78 & --14 21 24.9 &  1.18 & $   8.6 \pm    4.1$ & S53\\
17248-28 & --                & 17 28 22.70 & --14 21 25.7 &  4.22 & $  17.1 \pm    5.7$ & S53\\
\hline
\end{tabular}
\end{center}
\end{minipage}
\end{table*}

\begin{table*}
\caption{{\em NuSTAR} Serendips with Optical or Near-IR Positions in L17\label{tab:serendips_oir}}
\begin{minipage}{\linewidth}
\begin{center}
\begin{tabular}{cccc} \hline \hline
         &                & Separation between    &                 \\
Serendip & {\em Chandra}  & {\em Chandra} and L17 & L17             \\
ID       & Sources        & (arcseconds)\footnote{The 90\% confidence uncertainty on the position, including statistical and systematic contributions.}          & Classification\\ \hline
\multicolumn{4}{c}{Confirmations}\\ \hline
P347     & 17246-1  & $0.38\pm 0.73$ & AGN\\ \hline
P388     & 17248-1  & $0.29\pm 0.74$ & AGN\\
         & 17247-19 & $0.78\pm 2.72$ & ''\\ \hline
P390     & 17247-4  & $0.31\pm 0.73$ & AGN\\ \hline
P391     & 17247-10 & $0.30\pm 1.15$ & AGN\\
         & 17248-15 & $0.76\pm 1.03$ & ''\\ \hline
\multicolumn{4}{c}{Non-Confirmations}\\ \hline
P392     & 17247-1  & $1.59\pm 0.80$ & AGN\\ \hline
P448     & 17245-1  & $2.23\pm 0.72$ & ?\\
         & 17245-2  & $24.93\pm 0.76$ & ''\\ \hline
P463     & 18089-1  & $24.37\pm 0.75$ & ?\\ \hline
S53      & 17247-12 & $4.10\pm 1.18$ & AGN\\
         & 17248-28 & $2.91\pm 4.22$ & ''\\ \hline
\end{tabular}
\end{center}
\end{minipage}
\end{table*}

\begin{table*}
\caption{Optical Emission Lines for P448\label{tab:lines}}
\begin{minipage}{\linewidth}
\begin{center}
\begin{tabular}{ccccc} \hline \hline
Line\footnote{These are from the Keck/LRIS measurements of IPHAS~J202421.67+335050.1, which is the optical counterpart of CXOU~J202421.6+335050.}     & Wavelength  & EW\footnote{The equivalent width (EW) and the flux values are measured before dereddening.}          & FWHM\footnote{Corrected for the instrumental resolution of 215\,km/s at H$\alpha$ and 248\,km/s at H$\beta$.} & Flux$^{b}$\\
 & (Angstroms) & (Angstroms) & (km/s) & ($10^{-17}$\,erg\,cm$^{-2}$\,s$^{-1}$)\\ \hline
H$\beta$  & $4859\pm 0.6$ & $14.2\pm 1.3$ & $949\pm 43$ & $7.0\pm 0.9$\\
H$\alpha$ & $6561\pm 0.3$ & $20.1\pm 1.7$ & $1057\pm 14$ & $18.89\pm 0.08$\\ \hline
\end{tabular}
\end{center}
\end{minipage}
\end{table*}

\clearpage
\newpage

\begingroup
\renewcommand*{\arraystretch}{1.1}
\begin{longtable}{ccccccc}
\caption{{\em Chandra} Sources in {\em NuSTAR} serendip fields\label{tab:sourcelist}} \\
\hline 
\multicolumn{1}{c}{Source} & 
\multicolumn{1}{c}{$\theta$$^{a}$} & 
\multicolumn{1}{c}{{\em Chandra} R.A.} & 
\multicolumn{1}{c}{{\em Chandra} Decl.} & 
\multicolumn{1}{c}{Position}  & 
\multicolumn{1}{c}{ACIS} &  
\multicolumn{1}{c}{Other} \\
\multicolumn{1}{c}{Number} & 
\multicolumn{1}{c}{($^{\prime}$)} & 
\multicolumn{1}{c}{(J2000, h, m, s)}  & 
\multicolumn{1}{c}{(J2000, deg, $^{\prime}$, $^{\prime\prime}$)} & 
\multicolumn{1}{c}{Error$^{b}$ ($^{\prime\prime}$)} &
\multicolumn{1}{c}{Counts$^{c}$} & 
\multicolumn{1}{c}{Identification$^{d}$}\\ \hline
\endfirsthead
\caption{Continued}\\
\hline
\multicolumn{1}{c}{Source} & 
\multicolumn{1}{c}{$\theta$$^{a}$} & 
\multicolumn{1}{c}{{\em Chandra} R.A.} & 
\multicolumn{1}{c}{{\em Chandra} Decl.} & 
\multicolumn{1}{c}{Position}  & 
\multicolumn{1}{c}{ACIS} &  
\multicolumn{1}{c}{Other} \\
\multicolumn{1}{c}{Number} & 
\multicolumn{1}{c}{($^{\prime}$)} & 
\multicolumn{1}{c}{(J2000, h, m, s)}  & 
\multicolumn{1}{c}{(J2000, $^{\circ}$, $^{\prime}$, $^{\prime\prime}$)} & 
\multicolumn{1}{c}{Error$^{b}$ ($^{\prime\prime}$)} &
\multicolumn{1}{c}{Counts$^{c}$} & 
\multicolumn{1}{c}{Identification$^{d}$}\\ \hline
\endhead
\hline
\endfoot
\hline
\endlastfoot
\multicolumn{7}{c}{ObsID 17246}\\ \hline
 1 &  0.30 & 14 54 40.60 & --51 35 15.1 &  0.73 & $  11.9 \pm    4.6$ & P347\\
 2 &  1.63 & 14 54 38.25 & --51 33 56.0 &  0.77 & $   9.9 \pm    4.3$ & --\\
 3 &  2.11 & 14 54 34.20 & --51 33 43.5 &  0.97 & $   3.8 \pm    3.2$ & --\\
 4 &  3.77 & 14 55 04.27 & --51 33 54.7 &  1.03 & $   8.8 \pm    4.1$ & --\\
 5 &  3.83 & 14 55 06.64 & --51 35 50.9 &  1.51 & $   3.9 \pm    3.2$ & --\\
 6 &  3.90 & 14 54 39.78 & --51 31 33.8 &  1.25 & $   5.7 \pm    3.6$ & --\\
 7 &  4.22 & 14 54 59.73 & --51 38 39.4 &  0.83 & $  30.9 \pm    6.6$ & --\\
 8 &  6.16 & 14 54 41.11 & --51 29 17.4 &  2.51 & $   6.2 \pm    3.8$ & --\\
 9 &  7.47 & 14 53 55.88 & --51 37 29.8 &  4.57 & $   5.4 \pm    3.6$ & --\\
10 &  7.98 & 14 53 52.74 & --51 37 40.4 &  3.41 & $   9.2 \pm    4.3$ & --\\
11 &  8.14 & 14 55 16.83 & --51 41 32.7 &  6.14 & $   5.1 \pm    3.6$ & --\\
12 &  8.71 & 14 54 02.47 & --51 29 17.3 &  4.06 & $   9.8 \pm    4.4$ & --\\
13 &  8.80 & 14 53 45.47 & --51 35 43.9 &  5.80 & $   6.8 \pm    4.0$ & --\\
14 & 12.29 & 14 54 03.27 & --51 24 44.8 &  2.38 & $  73.8 \pm    9.9$ & --\\
\hline
\multicolumn{7}{c}{ObsID 17247}\\ \hline
 1 &  0.33 & 17 28 05.76 & --14 21 08.2 &  0.83 & $   3.8 \pm    3.2$ & P392\\
 2 &  0.64 & 17 28 05.19 & --14 20 14.9 &  0.78 & $   5.8 \pm    3.6$ & P393\\
 3 &  0.72 & 17 28 07.15 & --14 20 24.6 &  0.77 & $   6.8 \pm    3.8$ & P393\\
 4 &  2.22 & 17 28 04.67 & --14 23 06.0 &  0.73 & $  33.8 \pm    6.9$ & P390\\
 5 &  2.44 & 17 28 06.64 & --14 18 28.9 &  0.85 & $   7.9 \pm    4.0$ & P394\\
 6 &  2.87 & 17 28 07.67 & --14 18 05.1 &  1.04 & $   4.9 \pm    3.4$ & P394\\
 7 &  3.51 & 17 27 51.15 & --14 21 58.4 &  1.39 & $   3.7 \pm    3.2$ & Gaia\\
 8 &  3.76 & 17 28 18.57 & --14 22 40.4 &  1.21 & $   5.7 \pm    3.6$ & --\\
 9 &  3.99 & 17 27 49.45 & --14 22 15.5 &  1.08 & $   8.7 \pm    4.1$ & Gaia\\
10 &  4.28 & 17 28 06.17 & --14 16 37.0 &  1.15 & $   8.8 \pm    4.1$ & Gaia, P391\\
11 &  4.29 & 17 28 02.96 & --14 25 08.5 &  1.42 & $   5.6 \pm    3.6$ & --\\
12 &  4.36 & 17 28 22.78 & --14 21 24.9 &  1.18 & $   8.6 \pm    4.1$ & S53\\
13 &  4.68 & 17 28 21.94 & --14 18 39.3 &  1.45 & $   6.8 \pm    3.8$ & --\\
14 &  4.86 & 17 27 59.90 & --14 16 10.7 &  1.33 & $   8.8 \pm    4.1$ & --\\
15 &  5.63 & 17 28 11.56 & --14 26 16.7 &  2.06 & $   6.3 \pm    3.8$ & --\\
16 &  6.12 & 17 28 19.78 & --14 15 56.0 &  0.73 & $1472.6 \pm   39.4$ & PDS~456\\
17 &  6.62 & 17 27 59.59 & --14 14 23.6 &  2.89 & $   6.4 \pm    3.8$ & --\\
18 &  6.64 & 17 27 54.47 & --14 27 01.4 &  2.77 & $   6.8 \pm    4.0$ & --\\
19 &  7.01 & 17 27 51.45 & --14 14 40.6 &  2.70 & $   8.3 \pm    4.1$ & P388\\
20 &  7.08 & 17 28 15.57 & --14 27 28.4 &  3.30 & $   6.6 \pm    4.0$ & --\\
21 &  9.48 & 17 27 46.00 & --14 29 10.9 &  2.24 & $  29.0 \pm    6.6$ & --\\
22 & 11.11 & 17 27 53.58 & --14 31 39.0 &  7.31 & $  11.2 \pm    5.0$ & --\\
23 & 11.49 & 17 28 06.94 & --14 09 24.1 &  2.15 & $  66.7 \pm    9.7$ & --\\
24 & 12.64 & 17 28 15.81 & --14 33 14.4 &  2.89 & $  59.1 \pm    9.1$ & --\\
\hline
\multicolumn{7}{c}{ObsID 17248}\\ \hline
 1 &  0.33 & 17 27 51.40 & --14 14 40.0 &  0.74 & $   9.8 \pm    4.3$ & P388\\
 2 &  2.13 & 17 27 59.49 & --14 14 23.8 &  0.73 & $  41.8 \pm    7.5$ & Gaia\\
 3 &  2.16 & 17 27 44.74 & --14 12 47.0 &  0.80 & $   9.8 \pm    4.3$ & --\\
 4 &  2.60 & 17 27 40.36 & --14 15 03.3 &  0.82 & $  10.8 \pm    4.4$ & Gaia\\
 5 &  2.72 & 17 28 01.95 & --14 14 24.4 &  1.02 & $   4.7 \pm    3.4$ & --\\
 6 &  2.80 & 17 28 02.03 & --14 14 55.5 &  0.97 & $   5.7 \pm    3.6$ & Gaia\\
 7 &  2.84 & 17 27 59.82 & --14 16 10.0 &  0.90 & $   7.7 \pm    4.0$ & --\\
 8 &  3.38 & 17 27 41.82 & --14 11 47.0 &  1.18 & $   4.8 \pm    3.4$ & --\\
 9 &  3.45 & 17 27 54.67 & --14 11 04.4 &  1.11 & $   5.8 \pm    3.6$ & --\\
10 &  3.49 & 17 27 37.98 & --14 12 45.9 &  1.01 & $   7.7 \pm    4.0$ & --\\
11 &  3.71 & 17 27 39.38 & --14 16 53.0 &  0.80 & $  28.6 \pm    6.4$ & Gaia\\
12 &  3.89 & 17 27 41.29 & --14 11 14.2 &  0.77 & $  53.8 \pm    8.4$ & --\\
13 &  4.03 & 17 28 06.86 & --14 13 25.4 &  1.47 & $   4.6 \pm    3.4$ & --\\
14 &  4.03 & 17 28 02.48 & --14 11 32.0 &  1.14 & $   7.8 \pm    4.0$ & --\\
15 &  4.35 & 17 28 06.12 & --14 16 37.3 &  1.05 & $  11.5 \pm    4.6$ & P391\\
16 &  5.03 & 17 27 40.36 & --14 10 01.4 &  1.81 & $   5.7 \pm    3.6$ & --\\
17 &  6.00 & 17 27 52.16 & --14 08 23.8 &  1.78 & $   9.4 \pm    4.3$ & --\\
18 &  6.30 & 17 28 06.64 & --14 09 24.9 &  0.93 & $  61.3 \pm    8.9$ & --\\
19 &  6.79 & 17 27 43.19 & --14 20 55.5 &  2.72 & $   7.4 \pm    4.1$ & --\\
20 &  7.21 & 17 28 19.76 & --14 15 56.6 &  0.76 & $ 928.1 \pm   31.5$ & PDS~456\\
21 &  7.85 & 17 27 49.45 & --14 22 14.0 &  6.02 & $   4.6 \pm    3.6$ & --\\
22 &  7.86 & 17 27 35.95 & --14 21 23.4 &  7.68 & $   3.6 \pm    3.4$ & --\\
23 &  8.54 & 17 27 42.03 & --14 06 06.8 &  2.85 & $  14.2 \pm    5.1$ & --\\
24 &  9.02 & 17 27 52.99 & --14 05 23.3 &  3.54 & $  12.9 \pm    5.0$ & --\\
25 &  9.13 & 17 28 23.85 & --14 18 44.1 &  4.58 & $   9.9 \pm    4.6$ & --\\
26 &  9.66 & 17 28 00.89 & --14 05 03.0 &  3.49 & $  16.4 \pm    5.4$ & --\\
27 & 10.29 & 17 28 30.65 & --14 17 53.6 &  5.54 & $  11.8 \pm    5.0$ & --\\
28 & 10.47 & 17 28 22.70 & --14 21 25.7 &  4.13 & $  17.5 \pm    5.7$ & S53\\
29 & 11.53 & 17 28 25.54 & --14 22 15.1 &  5.11 & $  18.9 \pm    6.0$ & --\\
\hline
\multicolumn{7}{c}{ObsID 18088}\\ \hline
 1 &  0.09 & 20 23 12.49 & +20 42 48.8 &  0.72 & $  18.7 \pm    5.4$ & P443\\
 2 &  0.31 & 20 23 13.78 & +20 42 45.3 &  0.72 & $  18.7 \pm    5.4$ & P443\\
 3 &  0.68 & 20 23 13.01 & +20 43 24.0 &  0.75 & $   8.6 \pm    4.1$ & Gaia\\
 4 &  1.75 & 20 23 19.68 & +20 42 16.1 &  0.88 & $   4.6 \pm    3.4$ & --\\
 5 &  1.85 & 20 23 07.24 & +20 41 20.1 &  0.89 & $   4.7 \pm    3.4$ & Gaia\\
 6 &  1.97 & 20 23 13.17 & +20 40 46.2 &  0.90 & $   4.6 \pm    3.4$ & --\\
 7 &  2.22 & 20 23 21.09 & +20 41 48.4 &  0.73 & $  34.6 \pm    7.0$ & --\\
 8 &  2.30 & 20 23 12.72 & +20 45 01.8 &  0.86 & $   6.8 \pm    3.8$ & --\\
 9 &  2.52 & 20 23 06.50 & +20 40 37.7 &  0.73 & $  50.6 \pm    8.2$ & WISE\\
10 &  2.60 & 20 23 22.50 & +20 43 51.0 &  0.95 & $   5.6 \pm    3.6$ & --\\
11 &  2.64 & 20 23 21.69 & +20 41 12.9 &  0.95 & $   5.6 \pm    3.6$ & --\\
12 &  2.86 & 20 23 21.89 & +20 44 32.6 &  1.18 & $   3.5 \pm    3.2$ & --\\
13 &  2.91 & 20 23 00.05 & +20 42 57.7 &  0.81 & $  14.6 \pm    5.0$ & --\\
14 &  2.98 & 20 23 02.39 & +20 40 53.9 &  1.21 & $   3.6 \pm    3.2$ & --\\
15 &  3.11 & 20 23 23.31 & +20 40 55.9 &  1.13 & $   4.5 \pm    3.4$ & --\\
16 &  3.14 & 20 23 08.38 & +20 45 42.8 &  0.89 & $   9.8 \pm    4.3$ & Gaia\\
17 &  4.23 & 20 23 15.65 & +20 46 53.7 &  1.14 & $   8.7 \pm    4.1$ & --\\
18 &  4.61 & 20 23 22.48 & +20 38 45.6 &  1.29 & $   8.1 \pm    4.1$ & --\\
19 &  5.24 & 20 23 23.26 & +20 38 08.4 &  1.71 & $   6.9 \pm    4.0$ & --\\
20 &  6.14 & 20 23 08.09 & +20 36 40.5 &  0.95 & $  49.0 \pm    8.2$ & --\\
21 &  6.78 & 20 23 04.91 & +20 49 16.7 &  2.59 & $   7.9 \pm    4.1$ & --\\
22 &  7.08 & 20 22 51.23 & +20 47 46.7 &  2.13 & $  11.7 \pm    4.7$ & --\\
23 &  7.48 & 20 23 00.47 & +20 49 39.5 &  2.43 & $  11.6 \pm    4.7$ & --\\
24 &  7.63 & 20 23 18.72 & +20 35 14.2 &  0.96 & $ 115.4 \pm   11.9$ & Gaia\\
25 &  7.87 & 20 23 18.04 & +20 50 29.0 &  1.58 & $  27.3 \pm    6.5$ & --\\
26 &  7.91 & 20 23 45.35 & +20 44 35.8 &  2.37 & $  14.3 \pm    5.1$ & --\\
27 &  8.27 & 20 22 49.82 & +20 49 04.8 &  2.68 & $  13.9 \pm    5.1$ & --\\
28 &  8.47 & 20 23 18.46 & +20 51 05.0 &  1.81 & $  27.8 \pm    6.6$ & --\\
29 &  9.50 & 20 23 50.97 & +20 45 46.5 &  1.90 & $  38.8 \pm    7.5$ & --\\
30 & 12.62 & 20 23 30.75 & +20 30 51.5 &  3.08 & $  53.0 \pm    9.0$ & --\\
31 & 15.20 & 20 23 07.27 & +20 57 52.5 &  3.57 & $  94.3 \pm   13.4$ & --\\
\hline
\multicolumn{7}{c}{ObsID 17245}\\ \hline
 1 &  0.14 & 20 24 21.69 & +33 50 50.3 &  0.72 & $  15.5 \pm    5.1$ & P448\\
 2 &  0.24 & 20 24 23.33 & +33 51 00.3 &  0.76 & $   7.6 \pm    4.0$ & P448\\
 3 &  0.84 & 20 24 18.23 & +33 50 48.8 &  0.78 & $   6.5 \pm    3.8$ & --\\
 4 &  1.47 & 20 24 27.65 & +33 51 51.7 &  0.75 & $  13.6 \pm    4.8$ & --\\
 5 &  1.88 & 20 24 23.64 & +33 49 04.0 &  0.83 & $   6.7 \pm    3.8$ & --\\
 6 &  2.63 & 20 24 27.22 & +33 48 30.6 &  0.86 & $   8.7 \pm    4.1$ & --\\
 7 &  2.99 & 20 24 23.31 & +33 53 54.4 &  0.81 & $  15.4 \pm    5.1$ & --\\
 8 &  3.21 & 20 24 06.77 & +33 50 47.9 &  1.19 & $   4.3 \pm    3.4$ & --\\
 9 &  3.53 & 20 24 07.81 & +33 52 47.3 &  0.97 & $   9.2 \pm    4.3$ & --\\
10 &  3.70 & 20 24 30.05 & +33 47 35.7 &  1.12 & $   6.6 \pm    3.8$ & --\\
11 &  3.98 & 20 24 03.83 & +33 52 02.0 &  0.71 & $ 714.1 \pm   27.8$ & V404~Cyg\\
12 &  4.34 & 20 24 41.47 & +33 49 13.8 &  1.18 & $   8.4 \pm    4.1$ & --\\
13 &  5.17 & 20 23 58.42 & +33 52 25.6 &  1.15 & $  14.5 \pm    5.1$ & --\\
14 &  5.61 & 20 24 09.83 & +33 55 54.7 &  1.53 & $  10.0 \pm    4.6$ & --\\
15 &  5.85 & 20 24 02.53 & +33 55 06.5 &  1.67 & $   9.7 \pm    4.6$ & --\\
16 &  6.92 & 20 23 53.11 & +33 54 18.0 &  1.05 & $  51.3 \pm    8.5$ & P445\\
17 &  6.95 & 20 24 42.95 & +33 45 27.7 &  2.28 & $  10.1 \pm    4.6$ & --\\
18 &  8.57 & 20 24 56.80 & +33 46 14.8 &  2.51 & $  17.1 \pm    5.7$ & --\\
19 &  8.68 & 20 23 59.22 & +33 43 40.5 &  3.61 & $  11.1 \pm    5.0$ & --\\
20 &  9.31 & 20 23 39.69 & +33 47 58.9 &  1.64 & $  48.1 \pm    8.3$ & P444\\
21 & 13.70 & 20 24 50.20 & +33 38 31.2 &  8.44 & $  19.5 \pm    9.3$ & --\\
22 & 16.89 & 20 25 07.79 & +33 36 56.1 &  5.16 & $  84.9 \pm   17.7$ & --\\
\hline
\multicolumn{7}{c}{ObsID 17704}\\ \hline
 1 &  2.88 & 20 24 03.23 & +33 49 03.7 &  1.10 & $   4.3 \pm    3.4$ & --\\
 2 &  2.88 & 20 24 13.09 & +33 54 25.6 &  1.07 & $   4.5 \pm    3.4$ & --\\
 3 &  2.89 & 20 23 56.11 & +33 53 29.9 &  0.89 & $   8.5 \pm    4.1$ & --\\
 4 &  3.46 & 20 24 02.59 & +33 55 07.3 &  0.95 & $   9.4 \pm    4.3$ & ObsID 17245\\
 5 &  3.62 & 20 24 00.39 & +33 48 29.3 &  0.78 & $  40.0 \pm    7.5$ & P446\\
 6 &  3.80 & 20 23 55.67 & +33 48 53.2 &  1.10 & $   7.3 \pm    4.0$ & --\\
 7 &  3.86 & 20 23 53.23 & +33 54 18.1 &  0.80 & $  34.3 \pm    7.0$ & P445\\
 8 &  4.14 & 20 24 09.92 & +33 55 54.4 &  1.14 & $   8.2 \pm    4.1$ & ObsID 17245\\
 9 &  4.34 & 20 24 19.01 & +33 55 25.0 &  1.73 & $   4.2 \pm    3.4$ & --\\
10 &  4.35 & 20 24 23.64 & +33 49 03.8 &  1.18 & $   8.6 \pm    4.3$ & ObsID 17245\\
11 &  5.73 & 20 24 24.80 & +33 56 15.1 &  1.47 & $  11.4 \pm    4.7$ & --\\
12 &  6.29 & 20 24 29.88 & +33 47 35.4 &  1.75 & $  11.1 \pm    5.0$ & ObsID 17245\\
13 &  6.89 & 20 23 39.67 & +33 48 00.8 &  1.25 & $  29.0 \pm    6.7$ & P444\\
14 &  7.00 & 20 23 51.52 & +33 57 58.6 &  2.49 & $   9.1 \pm    4.6$ & --\\
15 &  7.01 & 20 24 28.73 & +33 46 22.5 &  1.89 & $  13.7 \pm    5.5$ & --\\
16 &  7.02 & 20 23 39.70 & +33 47 47.2 &  1.13 & $  41.8 \pm    7.8$ & P444\\
17 &  7.51 & 20 24 41.37 & +33 49 13.7 &  4.16 & $   6.1 \pm    4.3$ & --\\
18 &  8.72 & 20 24 23.14 & +33 59 53.3 &  6.53 & $   5.9 \pm    4.6$ & --\\
19 &  9.05 & 20 23 54.72 & +33 43 09.0 &  4.34 & $  10.2 \pm    5.4$ & --\\
20 &  9.38 & 20 24 17.05 & +34 00 58.3 &  7.70 & $   6.2 \pm    4.9$ & --\\
21 & 10.37 & 20 24 56.45 & +33 49 50.1 &  8.49 & $   7.7 \pm    5.7$ & --\\
22 & 11.68 & 20 24 56.73 & +33 46 11.6 &  3.99 & $  27.0 \pm    8.1$ & --\\
23 & 13.25 & 20 23 05.06 & +33 48 58.7 &  7.15 & $  20.8 \pm   10.0$ & --\\
24 & 14.89 & 20 23 07.79 & +33 43 32.1 &  7.26 & $  31.4 \pm   13.0$ & --\\
25 & 15.83 & 20 25 10.80 & +33 43 01.3 &  1.64 & $ 722.9 \pm   30.1$ & QSO~B2023+336\\
\hline
\multicolumn{7}{c}{ObsID 18089}\\ \hline
 1 &  0.31 & 21 19 35.52 & +33 36 44.0 &  0.75 & $   8.3 \pm    4.1$ & P463\\
 2 &  1.25 & 21 19 40.34 & +33 37 31.5 &  0.91 & $   3.2 \pm    3.2$ & --\\
 3 &  1.36 & 21 19 38.41 & +33 35 52.6 &  0.79 & $   7.3 \pm    4.0$ & --\\
 4 &  1.57 & 21 19 27.30 & +33 37 01.0 &  0.93 & $   3.4 \pm    3.2$ & --\\
 5 &  1.60 & 21 19 30.27 & +33 38 18.1 &  0.84 & $   5.3 \pm    3.6$ & --\\
 6 &  2.30 & 21 19 35.41 & +33 34 42.8 &  0.85 & $   7.2 \pm    4.0$ & WISE\\
 7 &  2.44 & 21 19 24.40 & +33 35 54.1 &  0.77 & $  19.3 \pm    5.6$ & Gaia\\
 8 &  2.45 & 21 19 42.69 & +33 38 50.0 &  0.75 & $  24.6 \pm    6.1$ & Gaia\\
 9 &  2.53 & 21 19 34.26 & +33 39 32.2 &  0.76 & $  21.6 \pm    5.8$ & Gaia\\
10 &  2.60 & 21 19 36.87 & +33 34 26.9 &  0.85 & $   9.2 \pm    4.3$ & --\\
11 &  2.77 & 21 19 35.66 & +33 34 15.1 &  1.08 & $   4.2 \pm    3.4$ & --\\
12 &  3.07 & 21 19 38.99 & +33 34 03.7 &  1.07 & $   5.1 \pm    3.6$ & --\\
13 &  3.32 & 21 19 37.40 & +33 40 17.3 &  1.03 & $   6.6 \pm    3.8$ & --\\
14 &  3.44 & 21 19 18.36 & +33 36 50.9 &  0.80 & $  25.0 \pm    6.2$ & --\\
15 &  3.50 & 21 19 35.11 & +33 33 31.1 &  0.97 & $   8.9 \pm    4.3$ & --\\
16 &  3.59 & 21 19 26.19 & +33 33 54.8 &  0.79 & $  30.9 \pm    6.7$ & --\\
17 &  3.72 & 21 19 26.11 & +33 40 15.2 &  0.85 & $  18.5 \pm    5.4$ & Gaia\\
18 &  3.95 & 21 19 17.11 & +33 35 36.3 &  1.07 & $   8.7 \pm    4.3$ & --\\
19 &  4.23 & 21 19 29.13 & +33 32 56.9 &  0.70 & $5212.5 \pm   73.2$ & Gaia, LEDA~2034356\\
20 &  4.39 & 21 19 25.18 & +33 33 06.8 &  1.10 & $  10.4 \pm    4.6$ & --\\
21 &  4.75 & 21 19 16.88 & +33 34 05.1 &  1.26 & $   9.2 \pm    4.4$ & --\\
22 &  4.80 & 21 19 20.72 & +33 33 13.5 &  1.28 & $   9.2 \pm    4.4$ & --\\
23 &  5.35 & 21 19 45.09 & +33 32 06.2 &  1.54 & $   8.6 \pm    4.4$ & --\\
24 &  5.44 & 21 19 18.99 & +33 32 41.2 &  1.61 & $   8.4 \pm    4.4$ & Gaia\\
25 &  5.94 & 21 19 52.83 & +33 41 37.8 &  1.86 & $   8.6 \pm    4.3$ & --\\
26 &  7.38 & 21 19 28.76 & +33 29 44.8 &  1.11 & $  53.1 \pm    8.5$ & WISE\\
27 &  7.87 & 21 19 49.72 & +33 44 15.1 &  2.43 & $  13.6 \pm    5.2$ & --\\
28 &  7.91 & 21 19 26.22 & +33 44 42.8 &  2.48 & $  13.4 \pm    5.2$ & --\\
29 &  7.97 & 21 19 56.19 & +33 43 37.5 &  1.90 & $  20.5 \pm    6.0$ & --\\
30 &  8.03 & 21 20 07.13 & +33 32 37.6 &  1.64 & $  27.4 \pm    6.6$ & --\\
31 &  8.27 & 21 19 46.35 & +33 44 56.0 &  2.34 & $  16.9 \pm    5.7$ & --\\
32 &  8.49 & 21 19 25.96 & +33 28 43.9 &  2.67 & $  15.2 \pm    5.6$ & --\\
33 &  8.54 & 21 19 51.59 & +33 44 48.8 &  5.65 & $   6.4 \pm    4.4$ & --\\
34 &  9.08 & 21 20 06.88 & +33 30 51.6 &  5.22 & $   8.4 \pm    4.9$ & --\\
35 &  9.44 & 21 20 16.71 & +33 33 23.8 &  2.39 & $  25.8 \pm    6.7$ & --\\
36 &  9.63 & 21 20 17.15 & +33 40 54.6 &  3.86 & $  14.4 \pm    5.7$ & --\\
37 &  9.64 & 21 19 53.64 & +33 45 49.1 &  3.21 & $  18.2 \pm    6.1$ & --\\
38 &  9.82 & 21 19 41.80 & +33 46 43.7 &  3.31 & $  18.6 \pm    6.2$ & --\\
39 & 10.03 & 21 19 58.01 & +33 45 48.8 &  2.15 & $  38.2 \pm    7.9$ & --\\
40 & 10.16 & 21 19 14.17 & +33 27 49.0 &  2.68 & $  28.1 \pm    7.2$ & --\\
41 & 10.53 & 21 19 47.83 & +33 47 11.8 &  2.79 & $  30.4 \pm    8.3$ & --\\
42 & 10.86 & 21 20 26.90 & +33 37 40.8 &  2.87 & $  32.5 \pm    7.7$ & --\\
43 & 11.09 & 21 20 17.76 & +33 30 26.9 &  4.13 & $  21.4 \pm    6.9$ & --\\
44 & 11.52 & 21 19 30.56 & +33 48 29.8 &  1.54 & $ 143.6 \pm   14.3$ & --\\
45 & 11.79 & 21 20 30.35 & +33 39 22.4 &  5.76 & $  17.8 \pm    6.8$ & --\\
46 & 12.39 & 21 20 21.43 & +33 29 18.7 &  4.08 & $  32.7 \pm    8.2$ & --\\
47 & 13.45 & 21 20 34.64 & +33 31 56.9 &  4.49 & $  39.3 \pm    9.2$ & --\\
\hline
\multicolumn{7}{c}{ObsID 18087}\\ \hline
 1 &  0.15 & 13 03 23.94 & --63 48 07.4 &  0.74 & $   9.7 \pm    4.3$ & --\\
 2 &  0.49 & 13 03 28.57 & --63 48 17.9 &  0.84 & $   3.7 \pm    3.2$ & --\\
 3 &  1.65 & 13 03 27.01 & --63 46 22.7 &  0.87 & $   4.7 \pm    3.4$ & VISTA\\
 4 &  1.81 & 13 03 08.61 & --63 47 49.7 &  0.88 & $   4.6 \pm    3.4$ & --\\
 5 &  2.26 & 13 03 13.96 & --63 49 55.7 &  0.90 & $   5.6 \pm    3.6$ & Gaia\\
 6 &  2.47 & 13 03 46.92 & --63 48 26.9 &  0.76 & $  20.7 \pm    5.7$ & Gaia\\
 7 &  2.50 & 13 03 12.60 & --63 45 55.4 &  0.93 & $   5.6 \pm    3.6$ & --\\
 8 &  2.58 & 13 03 05.10 & --63 49 23.6 &  0.88 & $   7.6 \pm    4.0$ & --\\
 9 &  2.66 & 13 03 22.75 & --63 45 22.1 &  0.79 & $  15.6 \pm    5.1$ & Gaia\\
10 &  2.66 & 13 03 01.26 & --63 47 29.4 &  1.02 & $   4.6 \pm    3.4$ & VISTA\\
11 &  3.90 & 13 02 50.04 & --63 47 23.1 &  1.20 & $   6.3 \pm    3.8$ & Gaia\\
12 &  4.63 & 13 02 47.63 & --63 50 08.5 &  0.72 & $ 479.1 \pm   22.9$ & Gaia, PSR~B1259--63\\
13 &  6.55 & 13 04 20.28 & --63 50 22.5 &  3.21 & $   5.5 \pm    3.8$ & --\\
14 &  6.65 & 13 03 51.08 & --63 54 00.4 &  1.65 & $  14.4 \pm    5.1$ & --\\
15 &  8.89 & 13 02 09.13 & --63 45 02.3 &  4.75 & $   8.7 \pm    4.7$ & --\\
16 & 10.39 & 13 02 12.71 & --63 54 42.1 &  6.01 & $  11.1 \pm    5.3$ & --\\
17 & 13.87 & 13 01 58.66 & --63 58 07.5 &  0.97 & $4084.0 \pm   65.1$ & S43\\
\hline
\end{longtable}
\endgroup
\begin{minipage}[b]{\textwidth}
$^{a}$The angular distance between the {\em Chandra} aimpoint and the source.\\ \\ $^{b}$The 90\% confidence uncertainty on the position, including statistical and systematic contributions.\\ \\ $^{c}$The number of ACIS counts detected (after background subtraction) in the 0.5--7 keV band (except for ObsID 17704, for which the energy band is 2--7\,keV).  The errors are 68\% confidence Poisson errors using the analytical approximations from \cite{gehrels86}.\\ \\ $^{d}$There is an entry in this column if the {\em Chandra} source may be identified with another source.  The identifications may be names of {\em NuSTAR} serendips, names of known sources, or names of catalogs if we simply identified the {\em Chandra} source with a source in an optical or near-IR catalog.
\end{minipage}

\begin{table*}
\caption{Position shifts\label{tab:shifts}}
\begin{minipage}{\linewidth}
\begin{center}
\begin{tabular}{cccc} \hline \hline
ObsID & Matches      & R.A. shift & Decl. shift\\
      &              & (arcsec)   & (arcsec)\\ \hline
17246 & --           & --         & --\\
17247 & 7            & 0.59 W     & 0.25 S\\
17248 & 4            & 0.10 E     & 0.36 S\\
18088 & 5            & 0.13 E     & 0.09 N\\
17245 & 1 (V404~Cyg) & --         & --\\
17704 & 4 (X-ray)    & 0.45 E     & 0.26 S\\
18089 & 9            & 0.12 E     & 0.08 S\\
18087 & 8            & 0.10 W     & 0.16 S\\ \hline
\end{tabular}
\end{center}
\end{minipage}
\end{table*}

\end{document}